\documentclass[twocolumn,apj,numberedappendix]{openjournal}

\usepackage{amsmath}
\usepackage[hidelinks]{hyperref}
\makeatletter
\renewcommand\frontmatter@makefntext[1]{\parindent 1em\noindent\@makefnmark #1}
\let\lastpage@putlabel\relax
\def\@eapj@cap@font{\normalfont}
\makeatother

\newcommand{\betapol}{\beta_{\mathrm{pol}}}

\newcounter{algo}

\shorttitle{SPT-3G Polarization Beam Characterization}
\shortauthors{SPT Collaboration}

\defcitealias{planck18-8}{Planck Collaboration et al. 2020a}
\defcitealias{planck18-6}{Planck Collaboration et al. 2020b}
\defcitealias{spt3gsoftware}{CMB-S4/SPT-3G Collaboration 2024}

\hypersetup{
  pdftitle={Characterization of the Polarization Beam Response of SPT-3G Using Point Sources},
  pdfauthor={T. de Haan et al.}
}

\begin{document}

\title{Characterization of the Polarization Beam Response of SPT-3G Using Point Sources}

\author{T.~de~Haan}
\affiliation{Institute of Particle and Nuclear Studies (IPNS), High Energy Accelerator Research Organization (KEK), 1-1 Oho, Tsukuba, Ibaraki 305-0801, Japan}
\affiliation{International Center for Quantum-field Measurement Systems for Studies of the Universe and Particles (WPI-QUP), High Energy Accelerator Research Organization (KEK), 1-1 Oho, Tsukuba, Ibaraki 305-0801, Japan}

\author{M.~Archipley}
\affiliation{Department of Astronomy and Astrophysics, University of Chicago, 5640 South Ellis Avenue, Chicago, IL, 60637, USA}
\affiliation{Kavli Institute for Cosmological Physics, University of Chicago, 5640 South Ellis Avenue, Chicago, IL, 60637, USA}

\author{N.~Huang}
\affiliation{Department of Physics, University of California, Berkeley, CA, 94720, USA}

\author{A.~J.~Anderson}
\affiliation{Fermi National Accelerator Laboratory, MS209, P.O. Box 500, Batavia, IL, 60510, USA}
\affiliation{Kavli Institute for Cosmological Physics, University of Chicago, 5640 South Ellis Avenue, Chicago, IL, 60637, USA}
\affiliation{Department of Astronomy and Astrophysics, University of Chicago, 5640 South Ellis Avenue, Chicago, IL, 60637, USA}

\author{B.~Ansarinejad}
\affiliation{School of Physics, University of Melbourne, Parkville, VIC 3010, Australia}

\author{L.~Balkenhol}
\affiliation{Sorbonne Universit\'e, CNRS, UMR 7095, Institut d'Astrophysique de Paris, 98 bis bd Arago, 75014 Paris, France}

\author{D.~R.~Barron}
\affiliation{Department of Physics and Astronomy, University of New Mexico, Albuquerque, NM, 87131, USA}

\author{P.~S.~Barry}
\affiliation{School of Physics and Astronomy, Cardiff University, Cardiff, CF24 3AA, UK}

\author{K.~Benabed}
\affiliation{Sorbonne Universit\'e, CNRS, UMR 7095, Institut d'Astrophysique de Paris, 98 bis bd Arago, 75014 Paris, France}

\author{A.~N.~Bender}
\affiliation{High-Energy Physics Division, Argonne National Laboratory, 9700 South Cass Avenue, Lemont, IL, 60439, USA}
\affiliation{Kavli Institute for Cosmological Physics, University of Chicago, 5640 South Ellis Avenue, Chicago, IL, 60637, USA}
\affiliation{Department of Astronomy and Astrophysics, University of Chicago, 5640 South Ellis Avenue, Chicago, IL, 60637, USA}

% ...

\begin{abstract}
Precise measurements of cosmic microwave background (CMB) polarization require rigorous control of instrumental systematics. For the South Pole Telescope's third-generation camera (SPT-3G), which observes in three frequency bands roughly centered on 95, 150, and 220~GHz, accurate characterization of the beam---the instrument's point spread function---is critical for understanding the polarized mm-wave sky. Here, we present direct measurements of SPT-3G's polarized beam response using observations of 100 bright extragalactic point sources.
Previous SPT-3G CMB power spectrum analyses introduced a phenomenological parameter $\betapol$ to describe the degree of polarization preserved in beam sidelobes. These analyses found evidence for significant depolarization driven by the requirement of polarization power spectrum consistency between different frequency bands.
Our direct measurements yield $\betapol = 0.89 \pm 0.10$ at 95~GHz, $1.08 \pm 0.10$ at 150~GHz, and $0.90 \pm 0.22$ at 220~GHz, indicating minimal sidelobe depolarization.
We validate these results through extensive systematic tests including Bayesian posterior sampling versus frequentist bootstrap resampling, real-space versus Fourier-space analysis, and variations on temperature-to-polarization leakage handling, covariance determination, and source selection. 
When compared to values inferred from previous cosmological analyses, which favored significant depolarization to resolve inter-frequency power spectrum inconsistencies, we find an effective difference of $1.3\sigma$. 
However, this apparent discrepancy is dependent on the beam modeling, as our point source-based analysis derives much of its constraining power on $\betapol$ from higher multipoles than the power spectrum analysis.
These measurements therefore admit three explanations for the frequency-dependent residuals observed in the power spectrum analysis: a statistical fluctuation, the need for more sophisticated polarized beam models, or systematics other than beam depolarization.
\end{abstract}



\section{Introduction}
\label{sec:intro}

Precise measurements of cosmic microwave background (CMB) polarization have emerged as powerful probes of fundamental physics, from constraining inflationary models through searches for primordial gravitational waves \citep{bicepkeck21c, tristram22, zebrowski25} to mapping the distribution of matter through gravitational lensing \citepalias[\citealp{qu24}]{planck18-8}. The South Pole Telescope's third-generation camera (SPT-3G) represents a major advance in CMB polarimetry, deploying approximately 16,000 transition-edge sensor (TES) bolometers observing at 95, 150, and 220~GHz \citep{benson14, sobrin22}. Realizing the scientific potential of this instrument requires rigorous control of systematic uncertainties, including the characterization of the polarized beam response.

The instrumental beam---the angular response to a point source---directly affects measurements of CMB power spectra through its multiplicative nature. Traditionally, the polarization response has been modeled by factorizing the temperature beam with detector-specific polarization properties, namely polarization efficiency and angle. This approximation is well motivated for the central, diffraction-limited main beam, but it need not hold exactly for the extended beam response. Part of the sidelobe response is the diffraction pattern expected from the finite optical aperture; the remainder is additional extended response arising from non-ideal effects such as scattering, reflections, mirror imperfections, gaps between primary mirror segments, surface roughness, and other optical processes. These components may have polarization properties that differ from those of the main beam, although we do not assume a priori that significant sidelobe depolarization should be present.

This possibility became relevant in previous SPT-3G power spectrum analyses, where \citet{ge24} found that the simple factorized beam model did not adequately describe the polarization spectra. In thermodynamic CMB temperature units, the CMB contribution to the $EE$ spectrum should be common across observing frequency. Disagreement among the 95, 150, and 220~GHz polarization spectra therefore motivates additional modeling of frequency-dependent foreground, beam, calibration, or other instrumental effects. \citet{ge24} introduced one phenomenological parameter $\betapol$ per frequency band to describe one such effect: a possible difference between the temperature and polarization beam response. In this model, $\betapol = 1$ corresponds to identical normalized temperature and polarization beams, while $\betapol = 0$ corresponds to the limiting case in which the extended sidelobe response contributes to temperature but not polarization.

The cosmological analysis of \citet{camphuis26} (hereafter C26) found strong evidence for $\betapol < 1$, with best-fit values under $\Lambda$CDM of $\betapol^{95}=0.48\pm0.13$, $\betapol^{150}=0.62\pm0.16$, and $\betapol^{220}=0.62\pm0.15$. The frequency-combined preference for $\betapol < 1.0$ had a significance exceeding $5\sigma$. This preference was driven primarily by the requirement of consistency between the 95, 150, and 220~GHz auto- and cross-spectra, largely independent of the assumed cosmological model (see Appendix B.2 of C26). Similar model-independent evidence was found in \citet{ge24}, where constraints on $\betapol$ were driven by the requirement of consistent fits to the $EE$ power spectra across multiple frequency bands. However, these constraints were derived indirectly by marginalizing over $\betapol$ within the power spectrum analysis itself. They could therefore, in principle, absorb any frequency-dependent residual that projects onto this beam template. A direct measurement of $\betapol$ from independent data is required to test the polarized-beam interpretation and to provide informative priors for future cosmological constraints.

Extragalactic radio sources provide useful probes of the polarized beam response. While limited by typically low polarization fractions ($\sim 2\%$), many are effectively point-like at SPT-3G's angular resolution ($\sim1'$) and sufficiently bright to constrain the polarization beam profile. By analyzing the radial profile of polarization signals around these sources, we can directly constrain how the polarization efficiency varies with angular distance from the beam center, measuring $\betapol$ independently from CMB power spectrum analyses.

In this work, we present the first direct measurements of $\betapol$ using 100 bright, polarized point sources observed by SPT-3G during the 2019--2023 observing seasons. Our analysis employs a simultaneous fitting framework that optimizes both beam parameters and source properties across multiple sources. We leverage modern computational tools including JAX for automatic differentiation and GPU acceleration to efficiently explore the high-dimensional parameter space. Extensive systematic tests validate the robustness of our measurements. 

This paper is organized as follows. 
Section~\ref{sec:observations} describes the SPT-3G observations and data processing, including the treatment of bolometer time constants. 
Section~\ref{sec:methodology} details our analysis methodology, covering beam parameterizations, the fitting procedure, and systematic tests. 
Section~\ref{sec:results} presents our measurements of $\betapol$ and alternative beam characterizations. 
Section~\ref{sec:discussion} discusses the physical interpretation and implications for cosmology, and we conclude in Section~\ref{sec:conclusions}.

\section{Observations and Data}
\label{sec:observations}

\subsection{SPT-3G Instrument and Survey}

The SPT-3G camera, installed on the 10-meter South Pole Telescope in austral summer 2016--2017, observes the sky with unprecedented sensitivity in three frequency bands centered at 95, 150, and 220~GHz \citep{benson14, sobrin22}. The focal plane contains approximately 16,000 TES bolometers configured for polarization-sensitive observations, with each pixel containing six detectors sensitive to the three observing bands and two orthogonal linear polarization states. The beam full width at half maximum (FWHM) is approximately 1.6, 1.2, and 1.0~arcminutes at 95, 150, and 220~GHz, respectively.

This analysis uses data from the 2019--2023 observing seasons, comprising both the primary 1500~deg$^2$ survey field (``SPT-3G Main'') and three additional fields observed during austral summer totaling 2650~deg$^2$ (``SPT-3G Summer''). The footprints of these fields are shown in previous SPT-3G publications \citep[e.g.][]{prabhu24journal}; we do not reproduce them here, as the survey geometry is not central to the beam-measurement method. The corresponding temperature white-noise map depths are $(3.0,\,2.5,\,8.9)\,\mu{\rm K}$-arcmin at $(95,\,150,\,220)\,$GHz for the main field, and $(9.8,\,9.2,\,29.2)\,\mu{\rm K}$-arcmin for the summer fields \citep{prabhu24journal}.

\subsection{Map Making and Data Processing}

The time-ordered data (TOD) from individual detectors are processed into maps using the standard SPT-3G pipeline \citepalias{spt3gsoftware}. 
The TOD are first filtered by fitting and subtracting polynomial and common-mode templates, as well as a basis of low-frequency Fourier modes. 
This filtering uses a 15~arcminute radius exclusion mask centered on the target point source, where data interior to the mask are excluded from template determination but the resulting templates are subtracted from all TOD. 
This masked filtering approach preserves the source emission at unity transfer function while partially removing contamination on scales larger than the $15'$ mask radius ($\ell\lesssim720$), including primary CMB fluctuations and atmospheric noise.
The $15'$ radius is a compromise: the mask must be large enough that the filtering does not significantly attenuate the beam response and sidelobes relevant to this analysis, but not so large that the baseline estimation loses too much uncontaminated data or excludes too many sources lying near the field edges.

Following filtering, the TOD are binned into Stokes $T$, $Q$, and $U$ maps using inverse-variance weighting based on detector noise properties. We produce custom $0.5^\circ \times 0.5^\circ$ maps at $0.1'$ pixel resolution approximately centered on each source position. Although $0.1'$ pixels oversample the approximately arcminute-scale beam, this fine pixelization reduces the error associated with evaluating a continuous, nonlinear beam model at pixel centers rather than integrating the model over finite pixel areas. Coarser pixels would likely still be adequate, but $0.1'$ pixels render this approximation negligible without introducing significant coverage problems or unnecessary computational cost.

\subsubsection{Treatment of Bolometer Time Constants}
\label{sec:tau_treatment}

Since the cosmological analysis of C26, we have changed our approach to handling bolometer time constants \citep[see e.g.][]{archipley25}. Bolometer time constants (typically $\sim$5~ms) introduce a phase delay in detector response that effectively broadens the beam along the scan direction. In the C26 analysis, these effects were absorbed into an effective beam model. For this work, and as we transition to this approach for future SPT-3G power spectrum analyses, we deconvolve time constant effects at the TOD level using measured time constant values for each detector, cleanly separating optical beam properties from detector response characteristics.

\subsubsection{Leakage Subtraction}
\label{sec:leakage}

Instrumental polarization---spurious polarization signals arising from temperature anisotropies---must be removed before analyzing the polarized beam. We construct empirical temperature-to-polarization leakage templates for each field through an iterative procedure. The templates are empirical, pixel-by-pixel $Q$ and $U$ patterns induced by a temperature point source; they are not restricted to a monopole, quadrupole, or any other prescribed angular structure.

The separation between leakage and intrinsic polarized source signal relies on the fact that the intrinsic source polarization is modulated by each source's polarization angle, rotating in the $Q$/$U$ plane from source to source, whereas the static instrumental leakage pattern is fixed in instrument/map coordinates by the temperature response. Averaging normalized maps over sources with different polarization angles therefore suppresses the intrinsic polarization and yields a first-order estimate of the static leakage pattern. Initially, we normalize each source's $Q$ and $U$ maps by its peak temperature flux density, then average these normalized maps across all sources with linear flux weighting to create ensemble-averaged leakage templates. These initial templates contain small residual contributions from true astrophysical polarization. To remove this contamination, we iterate: we fit the source and beam parameters using the framework described in Section~\ref{sec:fitting_framework}, subtract the best-fit polarized signal model from each source's $Q$ and $U$ maps, and then reconstruct the templates from these residual maps. 
The templates change negligibly after the first iteration; in the baseline analysis we perform five iterations and use the final templates for leakage subtraction.

\begin{figure}[t]
\centering
\includegraphics[width=\linewidth]{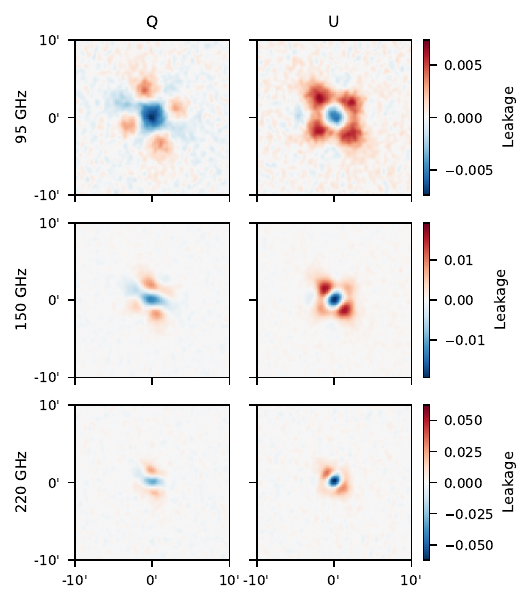}
\caption{Empirical temperature-to-polarization leakage templates for the main field. These templates represent the spurious $Q$ and $U$ signal induced by a temperature point source, constructed by averaging source maps after subtracting the best-fit astrophysical polarization models. The structure of these templates is consistent with the leakage patterns identified in the cosmological analysis \citep[see Figure 32 of][]{camphuis26}. These patterns are scaled by the source peak temperature and subtracted from each source prior to beam analysis.}
\label{fig:leakage_templates}
\end{figure}

Figure~\ref{fig:leakage_templates} shows the resulting leakage templates for the main survey field.
In addition to linear temperature flux weighting, we implement three alternative weighting schemes for template construction (median, flat, and quadratic) and explore these as systematic tests in Section~\ref{sec:systematic_tests}.

\subsection{Point Source Selection}

We identify the sample of point sources used for beam determination using a multi-step selection process. We begin with bright sources detected in SPT-3G temperature maps above a flux cutoff. For the main field, we choose a flux cutoff of $>200$~mJy at 95~GHz. For the summer fields, we choose a lower flux cutoff to ensure a statistically robust sample size from each field. This initial selection yields 129 candidate sources. From this set, we exclude 9 sources that appear extended or are associated with extended emission in visual inspection of the temperature map residuals (see the top-right panel of Figure~\ref{fig:source_example} for an example that passes this check).
We further exclude 11 sources which exhibit specific data quality issues, such as being located near the field boundaries where coverage is non-uniform. In such cases, the spatially varying noise levels would violate the stationarity assumptions required for our Fourier-domain analysis (Section~\ref{sec:chi2}). 
To facilitate a uniform multi-band analysis where the source position is determined with high signal-to-noise (S/N $\gg 10$) in every band, we exclude 3 sources with peak amplitudes $<300\,\mu\mathrm{K}$ (approximately 15\,mJy) in any of the three observing bands. Finally, we exclude 6 sources that have another bright source in the $0.5^\circ \times 0.5^\circ$ map, as that would contaminate the beam measurement. 

The final sample contains 100 sources distributed across our survey fields: 46 from the main field and 54 from the summer fields. All selected sources are used in the leakage-template construction (Section~\ref{sec:leakage}), independent of polarization fraction. We emphasize that the analysis is not based on an effectively unpolarized source sample: 91 of the 100 sources are detectably polarized, including all of the brightest sources, and the separation of leakage from intrinsic source polarization relies on the variation of polarization angles across the sample rather than on the assumption that the sources are unpolarized. The sources span a range of properties, with a few consistent with unpolarized emission and others up to $12\%$ polarized; the majority have polarization fractions in the 1\%--3\% range. The lower-polarization-significance sources still contribute to the leakage-template construction and to the empirical covariance estimation (Section~\ref{sec:precision}), while the $\betapol$ constraint is dominated by the significantly polarized sources (see Appendix~\ref{app:splitsample}).

\section{Methodology}
\label{sec:methodology}

\subsection{Beam Models}
\label{sec:beam_models}

Beam response functions can generally be described as two-dimensional or one-dimensional functions, expressible in either the harmonic or angular domain. In this work, we focus on the beam's one-dimensional radial profile $B(r)$, where $r$ represents the angular distance between the location of emission and the location of observation. We characterize this profile using two parameterizations: a regularized B-spline model (Section~\ref{sec:bspline}) with low model dependence and high parameter count, and the $\betapol$ model (Section~\ref{sec:betapol}) with strong physical motivation and low parameter count.

This choice to focus on the radial profile is motivated by the downstream application: for the SPT-3G power-spectrum analysis, the leading-order effect of the beam on the isotropized bandpowers is entirely captured by this radial profile. The full two-dimensional beam will be presented in H26; here we isolate the purely radial degree of freedom that controls the overall multiplicative suppression of $C_\ell$ in $T$, $E$, and $B$.
\\

\subsubsection{B-spline Models}
\label{sec:bspline}

We establish a flexible, data-driven beam model using B-splines. B-splines provide a general basis for representing smooth functions while maintaining computational efficiency for optimization.

During the development of this analysis, our initial implementation used a pure B-spline basis: a peak-normalized linear combination of cubic B-spline functions spanning the radial range $r \in [0, 10']$ with linearly spaced knots. However, this model proved susceptible to overfitting in the beam core. Depending on the value of the modeled source position relative to the nearest pixel center, the $B(r=0)=1$ knot in the model affected the innermost 1--4~pixels (6--24~arcseconds), where the true signal varies smoothly at sub-pixel scales. The optimizer could exploit the parametric freedom to uniformly raise all source flux parameters and lower all $r\ne0$ knots to create a small unphysical excursion in the beam core that fit noise fluctuations in these 1--4 pixels, improving $\chi^2$ at the cost of introducing spurious structure. This was not primarily a low-S/N-source effect; it arose from the interaction between source position, pixelization, and the peak-normalized spline basis in the highest-S/N part of the beam core.

Simply reducing the central knot density was considered but rejected: in practice it either left enough local freedom to reproduce these core excursions or made the transition between the core and the near sidelobes too rigid.

To regularize the beam core while maintaining flexibility at larger radii, we adopt a hybrid parameterization containing a central normalized Gaussian, plus B-splines for $r>0.25'$:
\begin{equation}
B(r) = \exp\left(-\frac{r^2}{2\sigma^2}\right) + \sum_{i=1}^{N} c_i \, \phi_i(r),
\end{equation}
where the Gaussian component (with free parameter $\sigma$) describes the central beam ($r \le 0.25'$) exclusively and the B-spline sum can modify this Gaussian at $r > 0.25'$. The B-spline basis functions $\{\phi_i(r)\}$ are constructed through a multi-step process to ensure proper regularization and numerical stability.

We begin with standard cubic B-splines ($k=4$) defined on a clamped knot vector with interior knots spaced by $\Delta r = 0.25'$ between $r_{\rm min} = 0.25'$ and $r_{\rm max} = 10'$. An earlier version of this analysis used a coarser interior knot spacing of $0.4'$, which proved too restrictive to describe the beam accurately; the finer $0.25'$ spacing adopted here changes the knot locations and improves the quality of the fits, modifying the fitted profiles shown in Figure~\ref{fig:bspline_beams} relative to that earlier version. Even at $0.25'$ spacing, the basis remains slightly too rigid to follow the steepest portion of the beam perfectly, which can produce a small negative excursion in the fitted profile near the transition from the main lobe to the first sidelobe; this limitation affects only the B-spline cross-check and not the $\betapol$-based primary results. To enforce smooth continuity with the Gaussian component, we impose the constraints $\sum_{i=1}^{N} c_i \, \phi_i(r=r_{\rm min}) = 0$ and $\sum_{i=1}^{N} c_i \, d\phi_i/dr|_{r=r_{\rm min}} = 0$. These constraints reduce the effective dimensionality of the B-spline parameter space.

We construct the constrained basis $\{\phi_i(r)\}$ by finding the null space of the constraint matrix via singular value decomposition. The resulting basis functions automatically satisfy the boundary conditions. To ensure numerical stability and interpretability, we orthonormalize these functions using 2D area normalization:
\begin{equation}
\int_0^\infty \phi_i(r) \, \phi_j(r) \, 2\pi r \, dr = \delta_{ij}.
\end{equation}
The orthonormalization is performed via eigendecomposition of the Gram matrix computed using the area-weighted inner product.

The final model has $1 + 2N$ free parameters, where $\sigma$ is shared between temperature and polarization beams, and $N=22$ orthonormal B-spline coefficients (determined by the number of knots minus the number of constraints) are fit independently for the temperature beam and polarization beam.

The resulting temperature beam fit is combined with data from Saturn observations (details in H26) to construct $B_{\mathrm{full}}(r)$ for the $\betapol$ model described in Equation~\ref{eq:betapol}, while the polarization beam fit (Figure~\ref{fig:bspline_beams}) provides an independent cross-check on that parameterization.

\subsubsection{\texorpdfstring{$\betapol$}{betapol} Model}
\label{sec:betapol}

Our primary beam model uses the $\betapol$ parameterization introduced in \citet{ge24}:
\begin{equation}
\label{eq:betapol}
B_{\mathrm{pol}}(r) = (1 - \betapol) B_{\mathrm{main}}(r) + \betapol B_{\mathrm{full}}(r),
\end{equation}
where $B_{\mathrm{pol}}(r)$ is the polarization beam profile we wish to constrain. This interpolates between two limiting cases: the main beam $B_{\mathrm{main}}(r)$, which describes only the central diffraction-limited response, and the full beam $B_{\mathrm{full}}(r)$, which includes both main beam and sidelobes as measured in temperature. The parameter $\betapol$ quantifies the polarization efficiency of the sidelobes, with $\betapol = 0$ indicating completely depolarized sidelobes and $\betapol = 1$ indicating identical temperature and polarization response at all radii.

The full beam $B_{\mathrm{full}}(r)$ combines the regularized B-spline temperature fit in the bright regime (beam response above $0.1\%$ of the peak) with data from dedicated Saturn observations in the faint regime (below $0.1\%$ of the peak). This leverages the strengths of both datasets: the highly linear point source measurements in the beam core and the high signal-to-noise Saturn measurements of the sidelobes.

The main beam $B_{\mathrm{main}}(r)$ is determined from a physical-optics model detailed in H26. This model accounts for the primary mirror, Lyot stop, and detector antenna patterns, parameterized by three free parameters that are fit simultaneously across all three observing bands using the innermost $0.75'$ of $B_{\mathrm{full}}(r)$. 
The main beam model predicts the diffraction-limited response of an idealized optical system.
While it naturally includes the near-sidelobe diffraction pattern, it contains no mechanism for producing anything close to the level of extended sidelobe power seen in the temperature data at $r \gtrsim 1'$.

The $\betapol$ model thus has only one free parameter per frequency band, in contrast to the $1 + 2N$ parameters per band in the regularized B-spline model. This parsimony makes $\betapol$ particularly valuable for cosmological analyses, where marginalizing over many beam parameters would significantly degrade cosmological constraints.

We interpret $\betapol$ as a sidelobe polarization amplitude parameter. Because it measures how closely the polarization beam resembles $B_{\mathrm{full}}(r)$ versus $B_{\mathrm{main}}(r)$, its value derived from the data depends on these two beam models. To explore this model dependence, we perform $\ell$-restricted $\betapol$ fits in Appendix~\ref{app:lrestricted}.

\subsection{Fitting Framework}
\label{sec:fitting_framework}

We determine beam parameters by simultaneously fitting the observed emission from all sources in our sample. For each source, we model the Stokes $T$, $Q$, and $U$ maps as the convolution of a point source with the instrumental beam, treating both the source properties and beam parameters as free parameters in a joint optimization.

\subsubsection{Model Parameterization}

For a source at fitted position $(x_0, y_0)$ with intrinsic Stokes parameters $(T_{\mathrm{src}}, Q_{\mathrm{src}}, U_{\mathrm{src}})$, the model prediction at map position $(x, y)$ is:
\begin{equation}
\begin{pmatrix} 
T_{\mathrm{model}}(x,y) \\ 
Q_{\mathrm{model}}(x,y) \\ 
U_{\mathrm{model}}(x,y) 
\end{pmatrix} = 
\begin{pmatrix} 
B_\mathrm{full}(r) \, T_\mathrm{src} \\
B_\mathrm{pol}(r) \, Q_\mathrm{src} \\
B_\mathrm{pol}(r) \, U_\mathrm{src}
\end{pmatrix},
\end{equation}
where $r = \sqrt{(x-x_0)^2 + (y-y_0)^2}$ is the angular distance from the fitted source position. The polarization beam $B_\mathrm{pol}(r)$ differs depending on the beam model. This formulation assumes azimuthally symmetric beams and neglects instrumental polarization (which has been removed via the leakage subtraction procedure described in Section~\ref{sec:leakage}).
We note that because the source Stokes amplitudes $(T_{\mathrm{src}}, Q_{\mathrm{src}}, U_{\mathrm{src}})$ are free parameters optimized for each source, our measurement of the beam's radial profile shape is insensitive to global calibration factors in temperature or polarization efficiency. Any global miscalibration is absorbed into the fitted source amplitudes, leaving the normalized beam profile constraints unaffected.

\subsubsection{Parameter Space}

The full parameter vector $\boldsymbol{\theta}$ contains both the beam parameters and the source parameters. The beam parameters depend on whether the chosen model is the $\betapol$ beam model or the regularized B-spline model, and whether the analysis is simultaneously fitting all bands or considering one band at a time.

For each point source, we fit five parameters per frequency band: the source positions $(x_0, y_0)$ and the Stokes amplitudes $(T_{\mathrm{src}}, Q_{\mathrm{src}}, U_{\mathrm{src}})$. For the full 100 sources, this yields 1500 source parameters that are optimized jointly with the beam parameters.

The total parameter count ranges from 501 (one-band, $\betapol$) to 1635 (three-band, regularized B-spline). This high-dimensional optimization is tractable due to the near-orthogonality of parameters, the presence of a single broad minimum, and the near-quadratic nature of that minimum. We leverage this structure through efficient gradient-based optimization as described in Section~\ref{sec:optimization}.

\subsubsection{Coordinate Systems and Implementation}

Source positions $(x_0, y_0)$ are expressed in pixel coordinates relative to each source's map center, with bounds of $\pm 5$ pixels ($\pm 0.5'$) from the nominal center. Source amplitudes are bounded to the range $(-5, 100)$~mK$_{\mathrm{CMB}}$ for both temperature and polarization. Both the position bounds and flux bounds were chosen to comfortably encompass the full range of observed source properties. Beam parameters are bounded to physically reasonable ranges. For $\betapol$, we allow the range $[-0.5, 2.0]$, B-spline coefficients are restricted to the range $[-0.5, 1.5]$, and the Gaussian width $\sigma$ is bounded to $[0.1, 2.0]$~arcminutes. These are numerical bounds for the optimization, not physical priors. While the physically interpretable range of the simple depolarization model is approximately $[0, 1]$, allowing a wider range avoids imposing artificial boundary effects and permits $\betapol$ to act as a diagnostic for model mismatch. None of the results in this work depend on the precise choices of these bounds: the posterior support lies far from the optimization boundaries, and the $\chi^2$ is extremely large everywhere on the hypercube boundary defined by these bounds.

Our bounds define a constrained optimization problem. To enable the use of efficient unconstrained gradient-based optimization algorithms (Section~\ref{sec:optimization}), we transform the bounded parameter space to an unconstrained space using a smooth bijective mapping. Specifically, we apply a scaled logit transformation:
\begin{equation}
\theta_{\mathrm{unconstrained}} = \log\left(\frac{\theta - \theta_{\mathrm{min}}}{\theta_{\mathrm{max}} - \theta}\right).
\end{equation}
This transformation maps the finite interval $(\theta_{\mathrm{min}}, \theta_{\mathrm{max}})$ to the real line $(-\infty, \infty)$, ensuring that the optimizer steps always yield valid physical parameters.

\subsection{Likelihood Formulation and Inference Methods}
\label{sec:statistics}

We evaluate beam and source parameters using maximum likelihood estimation, with uncertainties derived from bootstrap resampling. Our likelihood is based on a $\chi^2$ statistic that compares observed maps to model predictions. We implement this $\chi^2$ in both real space and Fourier space, and these implementations make different approximations about the noise properties and provide complementary systematic checks.

\subsubsection{\texorpdfstring{$\chi^2$}{Chi-squared} Formulations}
\label{sec:chi2}

Both formulations assume Gaussian noise with a known covariance structure. The real-space formulation operates directly on map pixels:
\begin{equation}
\chi^2_{\mathrm{real}} = \sum_{x,y,\nu,s,\nu',s'} \Delta_{x,y}^{\nu,s} \, W_{x,y}^{\nu,s,\nu',s'} \, \Delta_{x,y}^{\nu',s'},
\end{equation}
where $\Delta_{x,y}^{\nu,s} = d_{x,y}^{\nu,s} - m_{x,y}^{\nu,s}$ is the residual between data and model at pixel $(x,y)$ for frequency band $\nu$ and Stokes parameter $s \in \{T, Q, U\}$. The weight matrix $W_{x,y}^{\nu,s,\nu',s'}$ is the inverse covariance matrix (precision matrix) for that pixel, obtained from the map-making procedure. This matrix fully describes correlations between Stokes parameters within each band and pixel, but assumes independence between different pixels and between different bands. The pixel-independence assumption neglects the spatial correlations introduced by atmospheric noise and primary CMB fluctuations, but allows the weight matrix to vary spatially, capturing non-uniform noise properties across the map.

The Fourier-space formulation accounts for pixel-pixel correlations at the cost of assuming spatial stationarity:
\begin{equation}
\begin{split}
\chi^2_{\mathrm{Fourier}} = \sum_{\vec{k},\nu,s,\nu',s'} &\tilde{\Delta}_{\vec{k}}^{\nu,s*} \, (\mathbf{N}^{-1})_{\vec{k}}^{\nu,s,\nu',s'} \, \tilde{\Delta}_{\vec{k}}^{\nu',s'},
\end{split}
\end{equation}
where $\tilde{\Delta}_{\vec{k}}^{\nu,s} = \tilde{d}_{\vec{k}}^{\nu,s} - \tilde{m}_{\vec{k}}^{\nu,s}$ are the complex Fourier coefficients of the residuals at wavevector $\vec{k} = (k_x, k_y)$, and the asterisk denotes complex conjugation. The precision matrix $(\mathbf{N}^{-1})_{\vec{k}}$ describes correlations between bands and Stokes parameters at each Fourier mode, naturally incorporating pixel-pixel correlations. However, this formulation assumes the noise properties are uniform across the map.

The complementary approximations in these two formulations provide a valuable consistency check: agreement between real-space and Fourier-space results validates that neither set of assumptions introduces significant bias. Our baseline analysis uses the Fourier-space formulation with the precision matrix determination described below, as it more accurately captures the dominant correlated noise sources (atmospheric fluctuations and CMB). The real-space formulation serves as a systematic cross-check; despite our expectation that noise properties are highly uniform within the cutouts, this analysis validates that the stationarity assumption in the baseline analysis does not introduce bias.

\subsubsection{Fourier-Domain Precision Estimation}
\label{sec:precision}

The Fourier-space precision matrix $\mathbf{N}^{-1}$ is estimated from the input maps through an empirical procedure that accounts for both instrumental noise and the presence of primary CMB fluctuations.

First, we apply a noise mask to each input map. This mask consists of an apodization window at the map edges, combined with a circular excision around the source position. Both the map edges and the central circular cutout use a raised-cosine taper. The apodization suppresses edge effects in Fourier space, while the center excision removes the region dominated by the source signal, ensuring our noise estimate is not significantly contaminated by the signal we aim to measure.

We compute the discrete Fourier transform of each masked map and truncate the Fourier grid to modes with $|k_x| \leq k_{\mathrm{max}}$ and $|k_y| \leq k_{\mathrm{max}}$, with $k_\mathrm{max}$ corresponding to $\ell = 31{,}000$, which maps to 85\% of the Nyquist frequency of the TOD sampling, referred to sky coordinates at the highest-elevation, lowest-declination source. 
For each Fourier mode $\vec{k}$, we compute the empirical noise covariance matrix $\mathbf{N}_{\mathrm{emp}}$ by averaging over the $N_{\mathrm{src}}$ individual sources. The element corresponding to bands $\nu, \nu'$ and Stokes parameters $s, s'$ is calculated as:
\begin{equation}
(\mathbf{N}_{\mathrm{emp}})_{\vec{k}}^{\nu, s, \nu', s'} = \frac{1}{N_{\mathrm{src}}} \sum_{i=1}^{N_{\mathrm{src}}} \tilde{d}_{\vec{k}, i}^{\nu, s} \left( \tilde{d}_{\vec{k}, i}^{\nu', s'} \right)^*,
\end{equation}
where $\tilde{d}_{\vec{k}, i}^{\nu, s}$ is the Fourier coefficient of the masked map for source $i$, band $\nu$, and Stokes parameter $s$. This empirical covariance captures the average noise properties across the ensemble of observations. In order to account for differences in map depth, we group this procedure by observing field, with each field receiving a separate ensemble average.

To establish a noise floor, we estimate the noise level from the region ($|\ell_y| > 4000$ and $4000 < |\ell_x| < 8000$) where the noise is lowest and demonstrates little $k$-dependence. We take the 20th percentile of the variance in this region for each band and Stokes parameter and refer to this as the white noise floor. We clip all variances to be no lower than the white noise floor, which affects $\ll1\%$ of modes.

Our baseline covariance model combines theoretical CMB covariance with the empirical data covariance. The CMB covariance is computed from CAMB power spectra \citep{lewis99} using a Planck 2018 cosmology \citepalias{planck18-6}.
The angular power spectra $C_\ell^{TT}$, $C_\ell^{EE}$, $C_\ell^{BB}$, and $C_\ell^{TE}$ are transformed to the $(T, Q, U)$ basis via:
\begin{align}
C_{\vec{k}}^{T,T} &= C_\ell^{TT} \\
C_{\vec{k}}^{T,Q} &= C_\ell^{TE} \cos(2\phi), \quad C_{\vec{k}}^{T,U} = C_\ell^{TE} \sin(2\phi) \\
C_{\vec{k}}^{Q,Q} &= C_\ell^{EE} \cos^2(2\phi) + C_\ell^{BB} \sin^2(2\phi) \\
C_{\vec{k}}^{U,U} &= C_\ell^{EE} \sin^2(2\phi) + C_\ell^{BB} \cos^2(2\phi) \\
C_{\vec{k}}^{Q,U} &= (C_\ell^{EE} - C_\ell^{BB}) \sin(2\phi) \cos(2\phi)
\end{align}
where $\phi$ is the angle following the IAU polarization convention, and $|\vec{k}|$ maps to $\ell$ in the flat-sky approximation and our choice of $\vec{k}$ units. 
A raised-cosine high-pass filter is applied to the theoretical spectra to suppress fluctuations at $\ell \lesssim 720$, approximating the attenuation of modes larger than the mask size due to the TOD filtering.
The angular power spectra are then interpolated onto a Fourier grid of the map, supersampled by a factor of 4, then averaged down to the map resolution. The CMB covariance is then convolved with the apodization window function in the Fourier domain to account for mode coupling, and band-to-band covariances are constructed using band- and Stokes-dependent calibration factors from SPT-3G power spectrum analyses.

For diagonal elements of the covariance matrix, we take the element-wise maximum of the CMB-plus-white-noise model and the empirical covariance, ensuring we do not underestimate the noise. For off-diagonal elements, we use the CMB model. The elements of the correlation matrix are clipped to $[-0.8, 0.8]$ to ensure we do not rely on more than $5\times$ CMB suppression via band-band combinations. We find the resulting covariance matrix to be symmetric positive-definite for every mode. Finally, we invert each mode's covariance matrix to obtain the precision matrix $(\mathbf{N}^{-1})_{\vec{k}}$.

We also implement three alternative covariance models for cross-validation. The \textit{data-driven diagonal model} uses only the diagonal elements of the empirical covariance while neglecting band-band and Stokes-Stokes correlations. The \textit{full data-driven model} uses the entire empirical covariance matrix including all correlations. The \textit{white noise model} assumes uncorrelated noise across all non-zero modes, bands, and Stokes parameters, using only the measured white noise floor. 

\subsubsection{Bootstrap Uncertainty Estimation}
\label{sec:bootstrap}

Neither the real-space nor Fourier-space $\chi^2$ formulation perfectly captures the statistical properties of our data. A Bayesian exploration of the posterior assuming the noise model to be correct might therefore underestimate uncertainties. We employ bootstrap resampling to estimate parameter uncertainties. This bootstrap approach naturally accounts for potential contamination from marginally resolved sources, any mismatch between the assumed and true noise properties, source-to-source variations in noise properties, and effects such as the aforementioned pixel-pixel correlations in the real-space case or non-stationarity in the Fourier-space case.

We generate 200 bootstrap realizations by randomly resampling the 100 sources with replacement. For each realization, we re-optimize all parameters (both beam and source parameters) and record the resulting best-fit beam parameters. The sample covariance of these 200 bootstrap estimates provides our uncertainty estimate and parameter covariance matrix. 

\subsubsection{Optimization and Sampling}
\label{sec:optimization}

We explore the likelihood surface using both optimization (to find the maximum likelihood point) and Markov chain Monte Carlo sampling (to characterize the full posterior distribution). 

For optimization, we employ adaptive moment estimation algorithms. We use a two-stage approach: an initial phase with Adam \citep[Adaptive Moment Estimation,][]{kingmaAdamMethodStochastic2017a} followed by a refinement phase with AMSGrad \citep{reddiConvergenceAdam2019}, a variant that prevents the effective learning rate from increasing inappropriately when parameters become well-converged. This combination provides rapid and robust convergence across the high-dimensional parameter space.

For posterior sampling, we use microcanonical Langevin Monte Carlo \citep[MCLMC,][]{robnikMicrocanonicalHamiltonianMonte2023}, an efficient Hamiltonian Monte Carlo variant. We also validated results using the No-U-Turn Sampler \citep[NUTS,][]{hoffmanNoUTurnSamplerAdaptively2011}, finding excellent agreement but superior efficiency with MCLMC.

All optimization and sampling are performed in a transformed parameter space designed for efficiency. Starting from the bounded physical parameters, we first apply the logit transformation (Section~\ref{sec:fitting_framework}) to map to an unbounded space. We then scale each parameter by the inverse square root of its curvature (second derivative of $\chi^2$). This scaling makes the likelihood surface nearly isotropic, enabling efficient exploration. 

For MCMC sampling, we explicitly include the Jacobian of both transformations in the likelihood to ensure samples are drawn from the correct posterior distribution with an effective flat prior on the physical parameters. The near-diagonal curvature matrix (with typical off-diagonal correlations $<0.1$) validates our choice to scale by the diagonal curvature alone rather than computing the full Hessian, which would be too computationally expensive.

\subsubsection{Algorithmic Summary}
\label{sec:algorithm}

We summarize the baseline analysis pipeline in Algorithm~\ref{alg:baseline}. This procedure combines the iterative leakage removal, covariance estimation, and optimization procedures described in Sections~\ref{sec:observations}--\ref{sec:statistics}.

\begin{figure}[ht]
\refstepcounter{algo}
\label{alg:baseline} 
\begin{center}
\fbox{
\begin{minipage}{\dimexpr\linewidth-2\fboxsep-2\fboxrule\relax}
\small
\textbf{Algorithm \thealgo: Baseline Analysis} \\
\vspace{-0.5em}
\hrule
\vspace{0.5em}
\begin{tabbing}
\hspace{1.5em} \= \hspace{1em} \= \hspace{1em} \= \kill
\textbf{Require:} Sources $\mathcal{D} = \{ \mathbf{d}_i \}_{i=1}^N$ \\
\>  over bands $\nu \in \{95, 150, 220\}$; Stokes $s \in \{T, Q, U\}$ \\
\textbf{Require:} Beam templates $B_{\mathrm{main}}(r), B_{\mathrm{full}}(r)$; CMB $C_\ell$ \\
\\
\textbf{Subroutine: Joint Optimization} (Sec.~\ref{sec:optimization}) \\
\phantom{0}1: \> \textbf{function} \textsc{JointFit}($\mathcal{D}_{\mathrm{in}}$, $\mathbf{N}^{-1}$) \\
\phantom{0}2: \> \> Define $\boldsymbol{\theta} = [\betapol^{95}, \betapol^{150}, \betapol^{220}, \boldsymbol{\phi}_1^{95}, \dots, \boldsymbol{\phi}_N^{220}]$ \\
\phantom{0}3: \> \> \textbf{where} $\boldsymbol{\phi}_i^\nu = (x, y, T, Q, U)_i^\nu$ for source $i$, band $\nu$ \\
\phantom{0}4: \> \> Construct $B_{\mathrm{pol}}(r; \betapol^\nu)$ (Eq.~\ref{eq:betapol}) \\
\phantom{0}5: \> \> Generate model maps $m_i(\vec{x}; \boldsymbol{\phi}_i^\nu, B_{\mathrm{pol}}^\nu)$ \\
\phantom{0}6: \> \> $\tilde{\Delta}_{\vec{k}, i} \leftarrow \text{FFT}(\mathbf{d}_i - m_i)$ \\
\phantom{0}7: \> \> $\chi^2(\boldsymbol{\theta}) \leftarrow \sum_{i, \vec{k}} \tilde{\Delta}_{\vec{k}, i}^\dagger \mathbf{N}^{-1}_{\vec{k}} \tilde{\Delta}_{\vec{k}, i}$ \\
\phantom{0}8: \> \> Transform $\boldsymbol{\theta}$ to unbounded space \\
\phantom{0}9: \> \> $\boldsymbol{\theta}^* \leftarrow \text{Adam}(\nabla \chi^2, \boldsymbol{\theta}_{\mathrm{init}})$ \\
10: \> \> $\hat{\boldsymbol{\theta}} \leftarrow \text{AMSGrad}(\nabla \chi^2, \boldsymbol{\theta}^*)$ \\
11: \> \> \textbf{return} $\hat{\boldsymbol{\theta}}$ \\
\\
\textbf{Maximum Likelihood Analysis} (Sec.~\ref{sec:leakage}--\ref{sec:fitting_framework}) \\
12: \> Init $\mathbf{L}_s \leftarrow \text{WeightedMean}(\mathcal{D})_s$ for $s \in \{Q, U\}$ \\
13: \> \textbf{for} $j \leftarrow 1$ to $5$ \textbf{do} \\
14: \> \> $\mathcal{D}_{\mathrm{corr}} \leftarrow \{ \mathbf{d}_i - \mathbf{L} \}_{i=1}^N$ \\
15: \> \> $\mathbf{C}_{\mathrm{emp}} \leftarrow \text{Cov}(\text{Mask}(\mathcal{D}_{\mathrm{corr}}))$ \\
16: \> \> $\mathbf{C}_{\mathrm{CMB}} \leftarrow \text{SumWithModeCoupling}(C_\ell)$ \\
17: \> \> \textbf{for} each mode $\vec{k}$ \textbf{do} \\
18: \> \> \> $\mathbf{D}_{\vec{k}} \leftarrow \text{Max}(\text{diag}(\mathbf{C}_{\mathrm{emp}}(\vec{k})), \text{diag}(\mathbf{C}_{\mathrm{CMB}}(\vec{k})))$ \\
19: \> \> \> $\mathbf{N}^{-1}_{\vec{k}} \leftarrow (\mathbf{D}_{\vec{k}} + \text{OffDiag}(\mathbf{C}_{\mathrm{CMB}}(\vec{k})))^{-1}$ \\
20: \> \> $\hat{\boldsymbol{\theta}} \leftarrow \textsc{JointFit}(\mathcal{D}_{\mathrm{corr}}, \mathbf{N}^{-1})$ \\
21: \> \> $\mathbf{M}_i \leftarrow \text{Model}(\hat{\boldsymbol{\theta}}, i)$ \\
22: \> \> $\mathbf{L} \leftarrow \text{WeightedMean}(\{ \mathbf{d}_i - \mathbf{M}_i \}_{i=1}^N)$ \\
23: \> $\hat{\boldsymbol{\theta}}_{\mathrm{data}} \leftarrow \hat{\boldsymbol{\theta}}$ \\
\\
\textbf{Uncertainty Estimation} (Sec.~\ref{sec:bootstrap}) \\
24: \> $\Theta_{\mathrm{boot}} \leftarrow \emptyset$ \\
25: \> \textbf{for} $b \leftarrow 1$ to $N_{\mathrm{boot}}$ \textbf{do} \\
26: \> \> $\mathcal{D}^* \leftarrow \text{ResampleWithReplacement}(\mathcal{D}_{\mathrm{corr}})$ \\
27: \> \> $\hat{\boldsymbol{\theta}}^* \leftarrow \textsc{JointFit}(\mathcal{D}^*, \mathbf{N}^{-1})$ \\
28: \> \> $\Theta_{\mathrm{boot}} \leftarrow \Theta_{\mathrm{boot}} \cup \{ \hat{\betapol}^* \}$ \\
\\
\textbf{return} $\mu = \hat{\boldsymbol{\theta}}_\mathrm{data}$, $\Sigma = \mathrm{Cov}\left(\Theta_{\mathrm{boot}}\right)$
\end{tabbing}
\end{minipage}
}
\end{center}
\end{figure}

\subsection{Systematic Tests}
\label{sec:systematic_tests}

We perform extensive systematic tests to validate our baseline analysis and assess robustness to methodological choices. Our baseline uses the Fourier-space $\chi^2$ with CMB-informed precision matrices (Section~\ref{sec:precision}), bootstrap uncertainties (Section~\ref{sec:bootstrap}), and linear flux weighting for leakage templates. 

The systematic variations described below provide alternative analysis approaches that, while potentially having larger uncertainties, should yield unbiased estimates of the beam parameters. Appendix~\ref{app:postsub} documents additional pipeline-validation tests and their effect on $\betapol$.

\subsubsection{Temperature Beam Validation}

As a fundamental cross-check of our methodology, we apply the $\betapol$ interpolation framework to the temperature data. We fit for $\beta_T$ in the temperature beam:
\begin{equation}
B_{T}(r) = (1 - \beta_T) B_{\mathrm{main}}(r) + \beta_T B_{\mathrm{full}}(r),
\end{equation}
analogous to Equation~\ref{eq:betapol} for polarization. Since $B_{\mathrm{full}}(r)$ is constructed from the same temperature observations (supplemented by Saturn data), this analysis is circular by design. We therefore expect $\beta_T \approx 1.0$ with residuals driven only by the addition of Saturn data and numerical limitations, rather than statistical scatter. This provides an internal consistency check: if our fitting framework incorrectly recovers beam parameters, it should manifest equally in temperature and polarization.

We find $\beta_T^{95} = 1.000 \pm 0.025$, $\beta_T^{150} = 1.001 \pm 0.022$, and $\beta_T^{220} = 1.002 \pm 0.033$, which are all consistent with unity at $\lesssim0.05\sigma$. The high accuracy confirms that our methodology correctly recovers known beam properties. We note that this $0.05\sigma$ level should be interpreted as an internal pipeline-consistency scale for the tests discussed here, not as a complete beam-model uncertainty budget.

\subsubsection{Simulation Validation}

We validate the analysis pipeline through end-to-end simulations. We generate mock source maps with known input parameters: specified beam profiles (both $\betapol$ and Gaussian+B-spline models), source positions with $\sim0.1'$ offsets from map centers, and Stokes parameters spanning the range of observed source properties. The beam fitting analysis pipeline is then applied to these simulated sources. We find negligible systematic offsets, confirming that our implementation is free of significant coding errors or numerical artifacts and that the analysis framework is self-consistent.

\subsubsection{Posterior Distribution Characterization}

To validate our use of maximum-likelihood optimization with bootstrap uncertainties, we perform the MCLMC Bayesian analysis from Section~\ref{sec:optimization}. We generate 20,000 posterior samples after 5,000 adaptation steps to burn in and determine the trajectory length and step size. We use the resulting posterior samples to compute marginalized posterior distributions for all parameters.

Visual inspection of two-dimensional posterior slices for parameter pairs with the highest correlation coefficients shows that all posteriors are well-approximated by multivariate Gaussians with minimal off-diagonal correlations ($\rho \lesssim 0.3$). The marginalized beam parameter posteriors show good overlap with the bootstrap distributions, though they are not identical because the Bayesian posterior and frequentist bootstrap distributions answer fundamentally different statistical questions. The near-Gaussian posteriors validate our approximation that the likelihood surface is sufficiently well-behaved to justify maximum-likelihood estimation with bootstrap uncertainties.

\subsubsection{Real-Space Analysis}

Using the pixel-space $\chi^2$ (Section~\ref{sec:chi2}) with pixel-level precision matrices from map-making, we find consistent results with the baseline analysis. This analysis neglects pixel-pixel correlations but allows for spatially varying noise.

\subsubsection{Alternative Precision Matrices}

Using the three alternative precision matrices described in Section~\ref{sec:precision} (data-driven diagonal, fully data-driven, and white noise) in the Fourier-space $\chi^2$, we also find $\betapol$ constraints consistent with the baseline analysis. The white noise assumption in particular gives much larger uncertainties, though the results remain statistically consistent. This agreement validates our baseline precision matrix.

\subsubsection{Leakage Template Construction Variations}

The temperature-to-polarization leakage templates (Section~\ref{sec:leakage}) are constructed using linear flux weighting in our baseline analysis. We test three alternative weighting schemes: median (most robust to outliers), flat (least sensitive to source flux in the template), and quadratic (formally optimal in the case of unpolarized sources). We find that the $\betapol$ measurements are stable across all weighting choices, with variations $<0.2\sigma$ at all frequencies. This insensitivity demonstrates the effectiveness of the iterative template refinement in removing astrophysical polarization signal from the templates.

\subsubsection{Source Selection Robustness}
\label{sec:source_selection}

To test sensitivity to the source sample, we restrict the analysis to the 46 sources from the main field, excluding the 54 sources from the summer fields. This removes sources from the shallower summer-time observations. Table~\ref{tab:hyperparameters} shows that the resulting $\betapol$ constraints remain consistent with the full sample, though with slightly increased uncertainties due to the smaller sample size. 

\section{Results}
\label{sec:results}

\subsection{\texorpdfstring{$\betapol$}{betapol} Measurements}
\label{sec:betapol_results}

\begin{table*}[!htbp]
\centering
\begin{tabular}{llccc}
\hline\hline
Test Category & Variation & 95~GHz & 150~GHz & 220~GHz \\
\hline
\textbf{Baseline} & \textbf{---} & $\mathbf{0.890 \pm 0.098}$ & $\mathbf{1.084 \pm 0.095}$ & $\mathbf{0.896 \pm 0.219}$ \\
\hline
$\chi^2$ Space & Real-space & $0.719 \pm 0.301$ & $0.843 \pm 0.269$ & $0.787 \pm 0.358$ \\
\hline
Leakage Template & Median weighting & $0.928 \pm 0.097$ & $1.132 \pm 0.097$ & $0.788 \pm 0.220$ \\
 & Flat weighting & $0.911 \pm 0.118$ & $1.087 \pm 0.098$ & $0.906 \pm 0.224$ \\
 & Quadratic weighting & $0.867 \pm 0.089$ & $1.079 \pm 0.092$ & $0.919 \pm 0.217$ \\
\hline
Precision Matrix & White noise & $0.678 \pm 0.265$ & $0.793 \pm 0.211$ & $0.725 \pm 0.317$ \\
 & Full data-driven & $0.878 \pm 0.069$ & $1.020 \pm 0.089$ & $0.822 \pm 0.249$ \\
 & Data-driven diagonal & $0.852 \pm 0.085$ & $1.024 \pm 0.085$ & $0.753 \pm 0.233$ \\
\hline
Data Selection & 2019--2020, no $\tau$ decon & $0.833 \pm 0.167$ & $0.961 \pm 0.151$ & $0.605 \pm 0.352$ \\
 & Main field only & $0.855 \pm 0.106$ & $1.045 \pm 0.105$ & $0.828 \pm 0.249$ \\
\hline
Inference Method & Bayesian (MCLMC) & $0.914 \pm 0.114$ & $1.093 \pm 0.067$ & $0.866 \pm 0.171$ \\
\hline
\hline
\end{tabular}
\caption{Measurements of $\betapol$ under various analysis choices, demonstrating robustness of the baseline result to methodological variations. All systematic tests yield results consistent with the baseline within statistical uncertainties. Real-space analyses and white noise assumptions yield substantially larger uncertainties but remain consistent with the baseline. The tight agreement across leakage template constructions validates the iterative refinement procedure. The data-driven precision matrices demonstrate that the band-band-Stokes-Stokes correlations are of relatively minor importance to the overall $\betapol$ constraints. The Bayesian-frequentist comparison shows that these results are insensitive to the choice of statistical framework.}
\label{tab:hyperparameters}
\end{table*}

\begin{figure*}
\centering
\includegraphics[width=\linewidth]{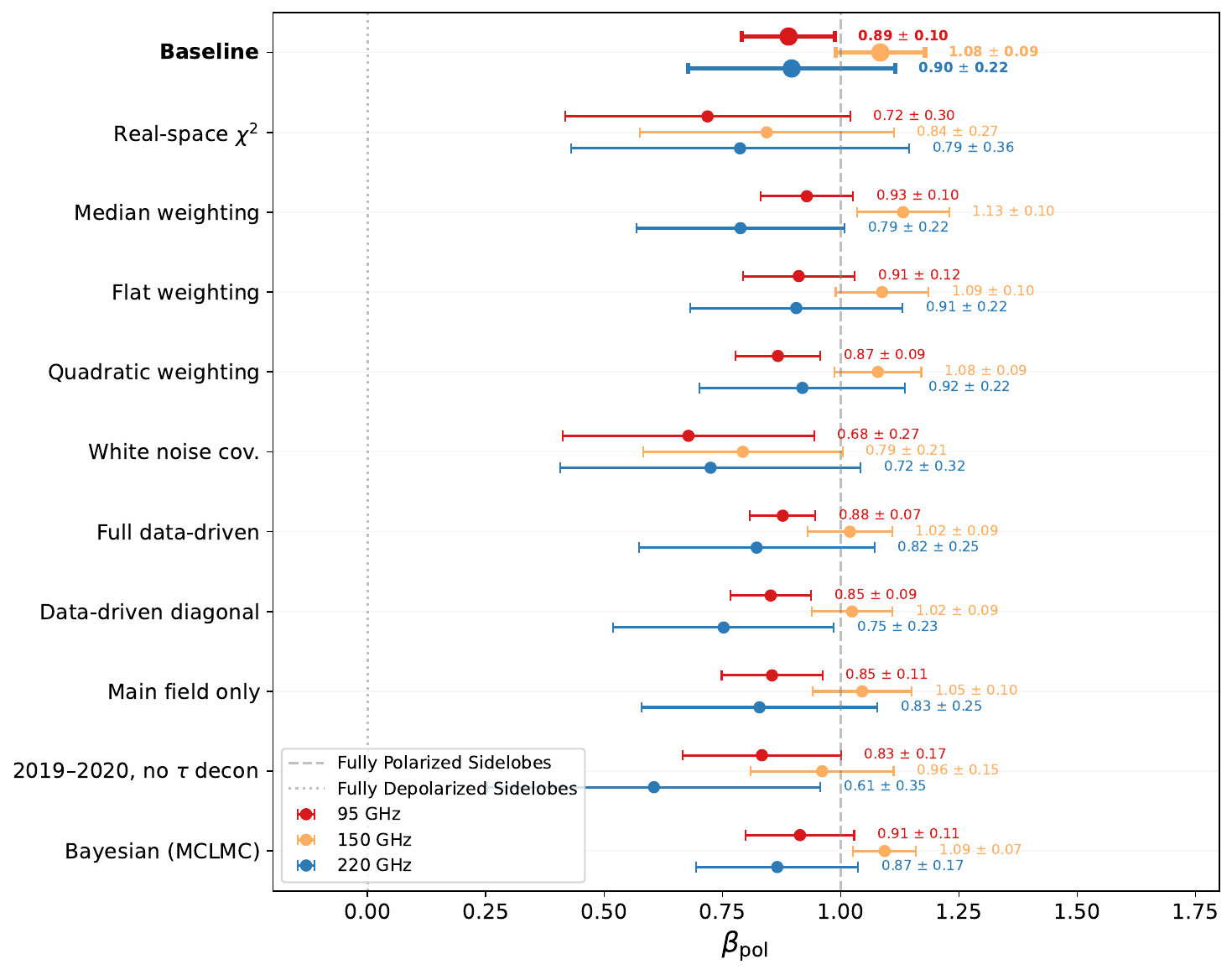}
\caption{Visualization of Table~\ref{tab:hyperparameters}. Points indicate the maximum-likelihood values (or posterior means for the Bayesian case), and error bars denote $1\sigma$ uncertainties (bootstrap or posterior standard deviation). The results are consistent across all variations, with the real-space and white-noise analyses yielding larger uncertainties as expected due to their suboptimal weighting of atmospheric noise. The tight agreement across leakage template methods and data subsets confirms that the measurement is not driven by specific analysis choices.}
\label{fig:hyperparameters}
\end{figure*}

Our primary measurements, along with systematic cross-checks, are summarized in Table~\ref{tab:hyperparameters} and visualized in Figure~\ref{fig:hyperparameters}. We find $\betapol = 0.89 \pm 0.10$ at 95~GHz, $1.08 \pm 0.10$ at 150~GHz, and $0.90 \pm 0.22$ at 220~GHz (baseline row). These values are consistent with $\betapol = 1$---indicating identical polarization and temperature beams, or $B^P=B^T$---at 1.1$\sigma$, 0.9$\sigma$, and 0.5$\sigma$ respectively. When considering the three bands jointly, the simultaneous measurement deviates from the $B^P=B^T$ case ($\betapol^{95}=\betapol^{150}=\betapol^{220}=1$) by a Mahalanobis distance of 1.5$\sigma$ (for three degrees of freedom; not directly comparable to the effective one-dimensional significances quoted elsewhere). This corresponds to a $p$-value of 0.54, indicating statistical consistency with $B^P=B^T$. The 220~GHz constraint is substantially weaker due to lower mapping speed and lower total polarized point source emission in this band.

\begin{figure*}
\centering
\includegraphics[width=\linewidth]{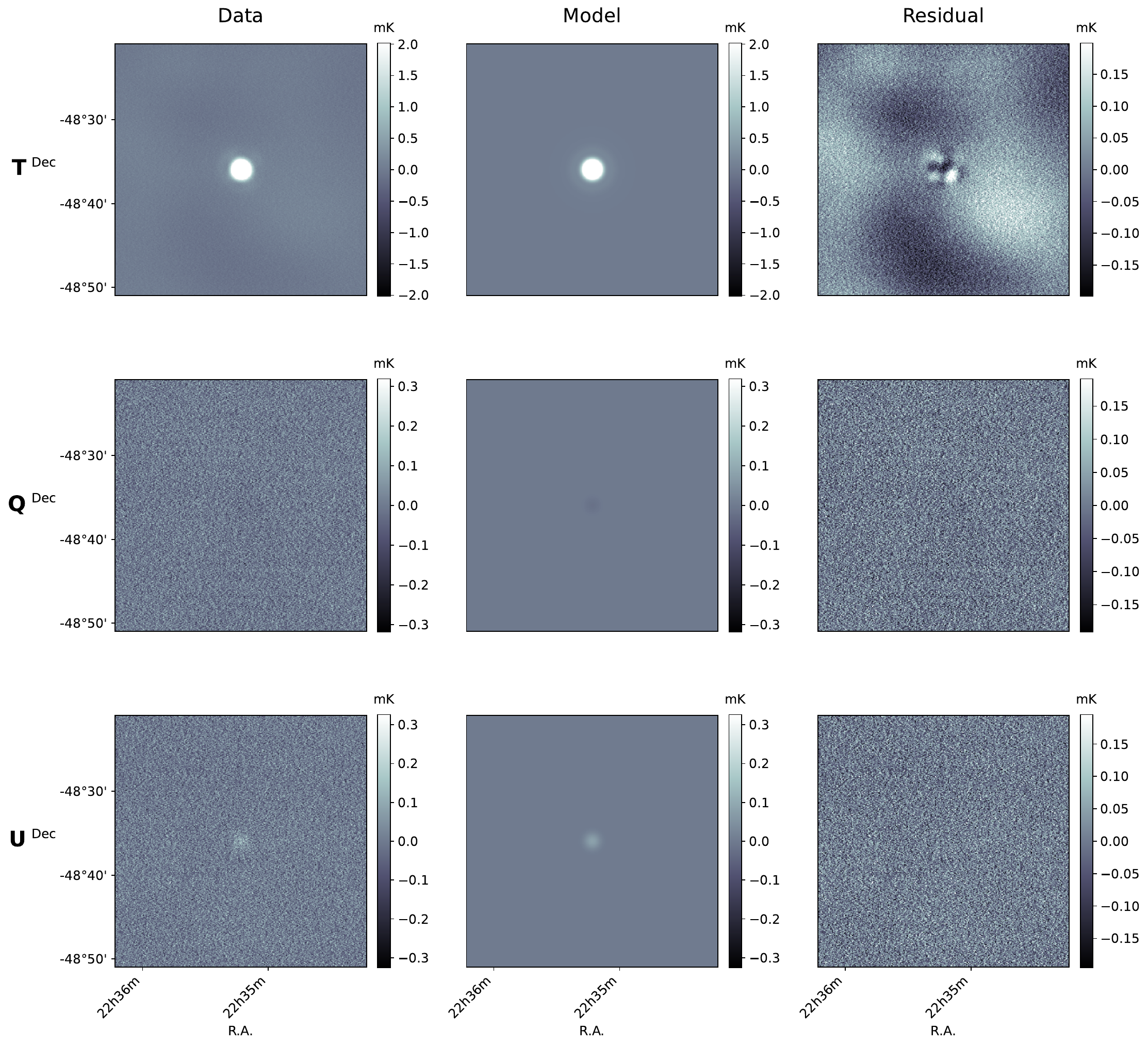}
\caption{Maps of a typical bright polarized point source in SPT-3G observations at 95~GHz. Rows indicate Stokes $T$, $Q$, and $U$ parameters. Columns display the observed data, best-fit $\betapol$ model, and residuals. The color scales are in units of $\mathrm{mK}_\mathrm{CMB}$, where one unit corresponds to the flux density needed to make the CMB appear 1~mK brighter at 95~GHz. The azimuthally symmetric model leaves residuals in Stokes $T$ due to the asymmetric nature of the beam. Stokes $Q$ and $U$ do not show obvious residual structure at the source location.}
\label{fig:source_example}
\end{figure*}

Figure~\ref{fig:source_example} illustrates the quality of our fits using a representative bright source at 95~GHz. The three rows display Stokes $T$, $Q$, and $U$ parameters, while columns show observed data, best-fit $\betapol$ model, and residuals. The model accurately reproduces the observed polarization pattern, with residuals consistent with noise expectations. The azimuthally symmetric beam model leaves structured residuals in Stokes $T$ due to the beam's true asymmetry. The corresponding asymmetry is not visible in Stokes $Q$ and $U$, but we caution that the absence of an obvious $Q$/$U$ residual asymmetry does not by itself prove the absence of an asymmetric polarized-beam component, since such a component would be suppressed by the low intrinsic polarization fraction ($\approx1.2\%$). Our argument that the polarized signal is separated from instrumental leakage instead rests on the leakage-subtraction procedure (Section~\ref{sec:leakage}) and the split-sample tests of Appendix~\ref{app:splitsample}. We note the limitation that any residual asymmetric polarized-beam component below our sensitivity would not be captured by the radial $\betapol$ model.

We validate the robustness of these measurements through extensive systematic tests (Section~\ref{sec:systematic_tests}), with results shown in the subsequent rows of Table~\ref{tab:hyperparameters}. 
For algorithmic choices that preserve the statistical power of the dataset, the results are stable to better than $0.2\sigma$. Methodological variations that significantly alter the estimator's precision yield central values that shift but remain statistically consistent with the baseline given their substantially larger uncertainties.

The real-space $\chi^2$ formulation yields results consistent with the baseline, though with significantly larger uncertainties (factor of $\sim$2--4 increase). This degradation in precision arises because the real-space covariance assumes pixel independence, failing to down-weight the noisy large-scale atmospheric and CMB fluctuations that dominate the variance. Two alternative precision matrix models in Fourier space show tighter agreement: the fully data-driven and diagonal data-driven models yield results within 0.2$\sigma$ of the baseline. The ``white noise'' Fourier analysis, which assumes a constant precision matrix (excluding the $\vec{k}=0$ mode), yields results and uncertainties comparable to the real-space analysis. The broad consistency between the baseline, real-space, alternative precision, and white-noise results confirms that our measurement is not driven by specific details of the noise modeling, although accurate covariance estimation is required to achieve high precision.

Leakage template construction proves remarkably stable, with all four weighting schemes (median, flat, linear, quadratic) yielding measurements within 0.2$\sigma$ of each other at all bands. This insensitivity validates the iterative template refinement procedure, confirming that residual astrophysical polarization signal has been effectively removed from the templates. Similarly, restricting the analysis to the 46 sources from the main field---excluding the 54 shallower summer field sources---yields consistent results with modestly increased uncertainties, demonstrating that no particular subset of sources drives the measurement.

The agreement between frequentist (bootstrap) and Bayesian (MCLMC posterior sampling) approaches further validates our uncertainty estimation. The Bayesian credible intervals are consistent with bootstrap confidence intervals at all frequencies. This consistency confirms that the likelihood surface is well-approximated by a multivariate Gaussian, justifying our use of maximum-likelihood estimation with bootstrap resampling.

\subsection{Source of \texorpdfstring{$\betapol$}{betapol} Precision}
\label{sec:precision_source}

To understand which parts of the data drive the $\betapol$ constraints and to inform future analyses, we decompose the $\chi^2$ curvature with respect to $\betapol$ (conditional likelihood) by computing partial sums across sources, Fourier modes, and Stokes parameters.

Stokes $Q$ and $U$ contribute asymmetric constraining power, with $Q$ providing approximately 60\% and $U$ providing 40\% of the total. 
This asymmetry arises from the azimuthal structure of the precision matrix. Modes approximately aligned with the $k_x$ and $k_y$ axes suffer contamination from low-$\ell$ fluctuations coupled in by the square window function. For Stokes $Q$, primary CMB $E$-mode power further degrades these modes, leaving the diagonal directions (along $k_x \pm k_y$) relatively clean and highly informative. In Stokes $U$, the noise contributions from mask coupling and CMB $E$-mode power affect complementary mode orientations, resulting in more uniform but overall lower precision when summed across the Fourier plane.

The radial $\ell$-dependence reveals that, combining the constraining power across bands, intermediate angular scales dominate the total, though the precise $\ell$ range that contributes most is not identical in every band. The monopole ($\ell=0$) contributes $<$1\% of the total curvature; source-to-source flux variations show this mode to be too noisy to constrain the subtle effects of $\betapol$. Low-$\ell$ modes ($\ell < 3000$) contribute approximately 45\% of the total precision on $\betapol^{95}$ and $\betapol^{220}$, and $\sim$30\% on $\betapol^{150}$. The cumulative curvature reaches $>$90\% of its asymptotic value by $\ell = 13{,}000$ across all three bands, indicating that modes above $\ell \sim 15{,}000$ add minimal constraining power on $\betapol$.

Individual sources contribute heterogeneously but without extreme outliers. The most constraining source (SPT-S J025329-5441.8 at 150~GHz) provides 12\% of the total precision, while no other single source exceeds 7\%. For comparison, uniform contribution across 100 sources would yield 1\% per source. This moderate heteroskedasticity validates our methodological choices of the source-averaged construction of leakage templates and the use of bootstrap resampling for uncertainty estimation. To verify directly that the measurement is not driven by a single source or by a particular source subset, we repeat the analysis on independent source subsets in Appendix~\ref{app:splitsample}; the resulting $\betapol$ constraints are statistically consistent with the baseline. We nonetheless retain the full-sample baseline because it is statistically optimal under the model and because the bootstrap resampling already accounts for source-to-source variation.

\subsection{B-spline Beam Profiles}

\begin{figure*}
\centering
\includegraphics[width=\linewidth]{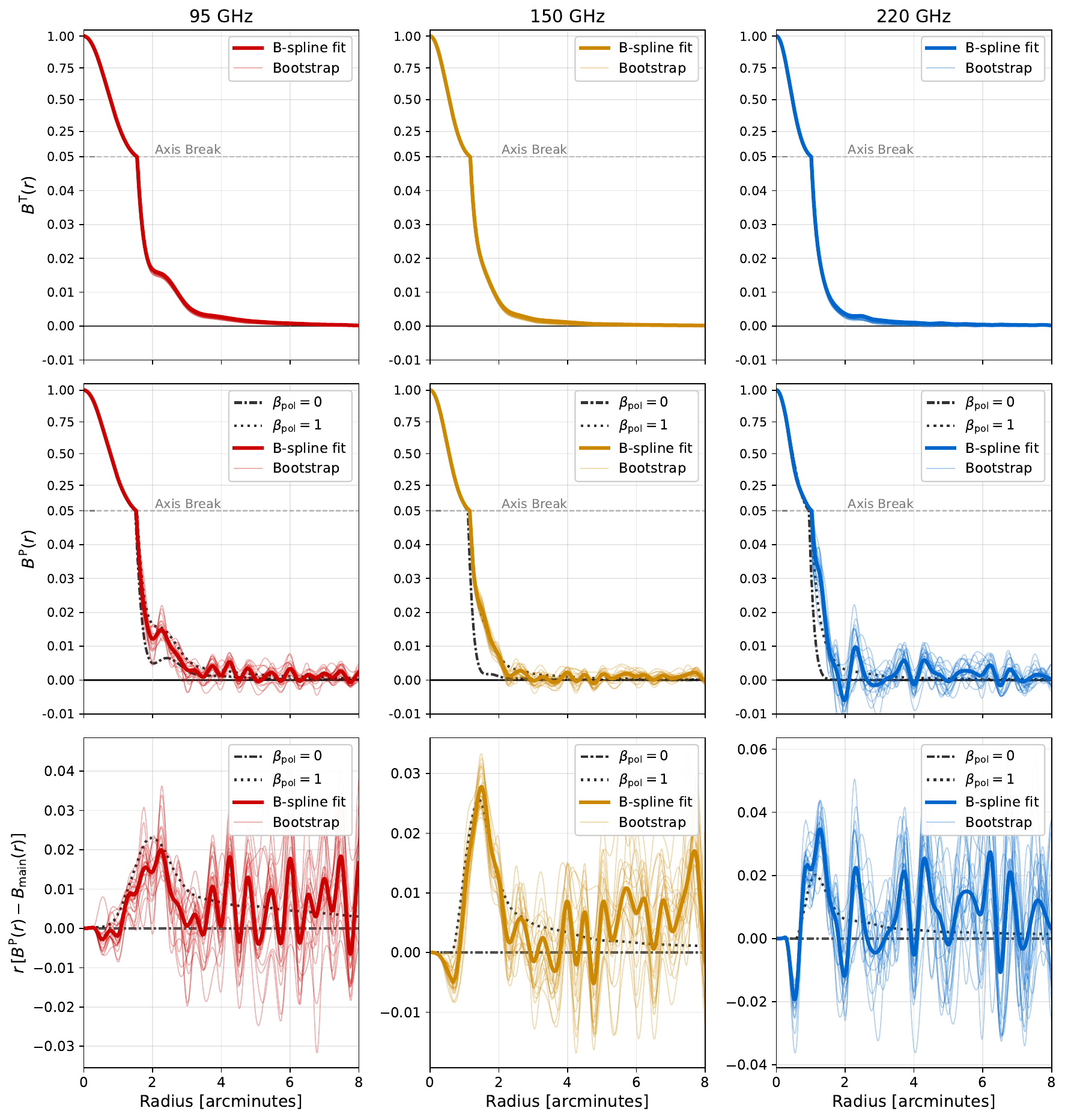}
\caption{Temperature (top) and polarization (middle and bottom) radial beam profiles reconstructed using the B-spline basis. These profiles are peak-normalized such that $B(r=0)=1$. Thick colored curves show the best-fit profiles for 95~GHz (red), 150~GHz (gold), and 220~GHz (blue). Thin colored lines show 20 bootstrap realizations, illustrating the measurement uncertainty. The polarization beam uncertainties are approximately 50~times larger than the temperature beam uncertainties due to the $\sim2\%$ intrinsic polarization fraction of most sources. Overlaid on the polarization beams are predictions from the $\betapol$ model for two limiting cases: fully depolarized sidelobes ($\betapol=0$, dot-dashed black) and identical polarization and temperature beams ($\betapol=1$, dotted black). The lower row of panels shows the residuals of the polarization profiles relative to the $\betapol=0$ prediction, so that the evidence for sidelobe polarization can be assessed as a function of radius. The model-independent B-spline fits broadly track the $\betapol=1$ prediction and show no statistically significant evidence for sidelobe depolarization.}
\label{fig:bspline_beams}
\end{figure*}

\begin{figure*}
\centering
\includegraphics[width=\linewidth]{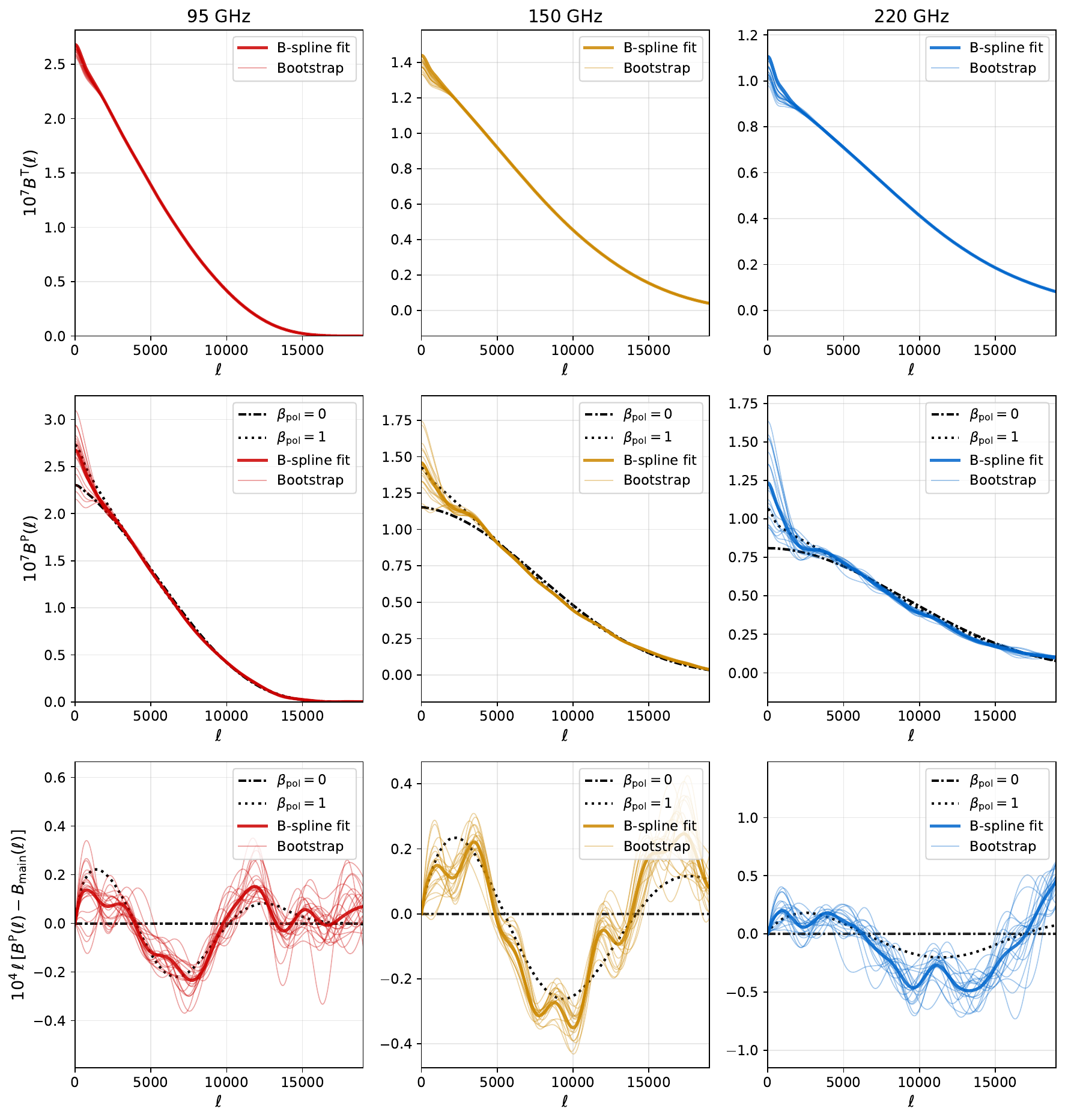}
\caption{Harmonic-domain representation of the beam profiles shown in Figure~\ref{fig:bspline_beams}. The curves are obtained via Hankel transformation of the real-space B-spline fits and contain equivalent information. The profiles retain the real-space peak normalization $B(r=0)=1$, corresponding to $\int \ell B(\ell) d\ell = 2\pi$. The lower row of panels shows the residuals of the measured polarization profiles relative to the $\betapol=0$ prediction, plotted as $\ell \left[ B^\mathrm{P}(\ell)-B_\mathrm{main}(\ell) \right]$. These residuals allow for a qualitative assessment of the evidence for the $\betapol=1$ model as a function of $\ell$.}
\label{fig:bspline_beams_harmonic}
\end{figure*}

To assess the qualitative consistency of the $\betapol$ model with the polarized beam measurements, we fit the highly parametric regularized B-spline model defined in Section~\ref{sec:bspline} and examine the resulting radial beam profiles. The B-spline model provides a useful cross-check because, unlike the $\betapol$ parameterization, it imposes no fixed relationship between the temperature and polarization beam shapes beyond the central $0.25'$.

Figure~\ref{fig:bspline_beams} displays the reconstructed real-space profiles. The temperature beams are characterized with high precision across the full radial range. The polarization beams exhibit fractional uncertainties approximately 50~times larger, a direct consequence of the low intrinsic polarization fractions ($\sim 2\%$) of the extragalactic source population. Despite these larger uncertainties, the reconstructed polarization profiles are broadly consistent with the $\betapol = 1$ case within their uncertainties at all three frequencies.

The flexible B-spline profiles become noisy at larger radii, especially at 220~GHz where the constraining power is weakest. While the B-spline basis has sufficient freedom to fit complex radial structure, the reconstruction does not reveal statistically significant deviations from the $\betapol = 1$ interpretation. Individual fluctuations in the best-fit profile should therefore not be overinterpreted as physical beam features. Likewise, the small negative undershoot visible near the transition from the main lobe to the first sidelobe, most prominently at 220~GHz, reflects the residual rigidity of the B-spline basis in this region of rapid variation (Section~\ref{sec:bspline}) rather than genuine beam structure; it is confined to this visualization and cross-check and does not enter the $\betapol$-based primary results. We conclude that the polarization beam measurements are consistent with the temperature beam within the statistical precision of the point-source data.

Figure~\ref{fig:bspline_beams_harmonic} presents the same reconstruction in the harmonic domain. The residual panels are plotted as $\ell$ times the difference between the reconstructed polarization beam and the $\betapol=0$ prediction. As detailed in the curvature analysis of Section~\ref{sec:precision_source}, modes with $\ell < 3000$ provide a substantial, but not dominant, fraction of the total constraining power: approximately 45\%, 30\%, and 45\% of the total precision at 95, 150, and 220~GHz, respectively. The remaining, comparable or larger, fraction of the information comes from $3000 \lesssim \ell \lesssim 10{,}000$, where the point-source data provide tighter constraints on the shape of $B(\ell)$. This distinction is important for interpreting the comparison to C26, which determined $\betapol$ using CMB power spectra over the lower multipole range $400 < \ell < 4000$.

While the $\betapol$ parameterization prescribes a specific scale dependence based on the temperature beam morphology and physical-optics modeling, the true polarized beam could follow a different profile. An alternative model could potentially reconcile these measurements by allowing the polarization response to deviate from the model prediction at low multipoles, while remaining consistent with the temperature beam over the higher multipoles probed by the point-source data. Under such a scenario, the apparent discrepancy between the C26 preference for depolarized sidelobes and the direct point-source measurement presented here would reflect the limitations of the single-parameter $\betapol$ model when applied to datasets sensitive to distinct angular regimes. We probe this scenario using $\ell_{\mathrm{max}}$-limited measurements in Appendix~\ref{app:lrestricted}.

\subsection{Comparison with Previous Constraints}
\label{sec:cosmology_comparison}

\begin{figure*}
\centering
\includegraphics[width=\linewidth]{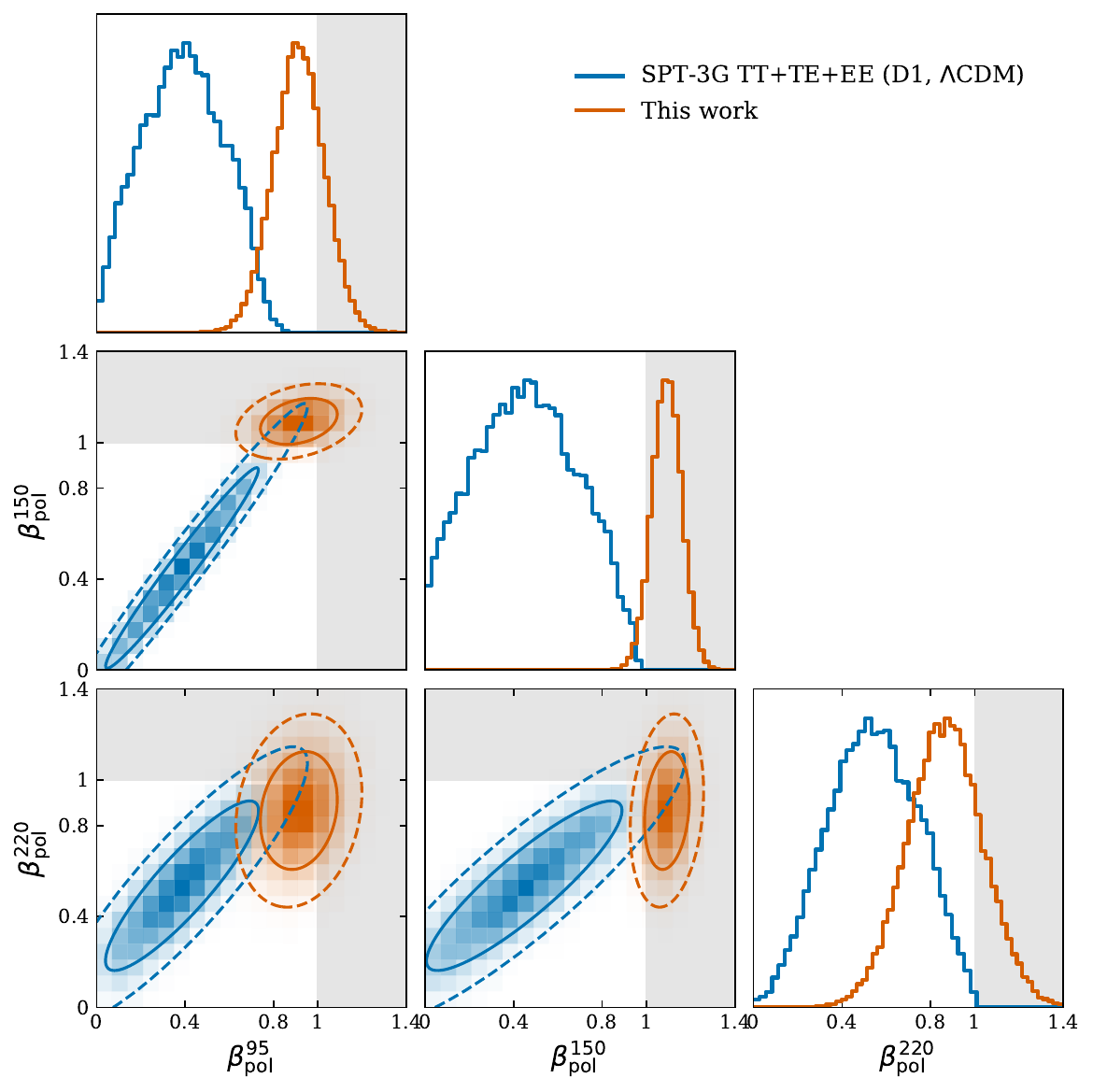}
\caption{Comparison of $\betapol$ measurements from this work (orange; Bayesian results from Table~\ref{tab:hyperparameters}) with values inferred from the C26 cosmological power spectrum analysis (blue). The 1D histograms and color density grids display the distribution of MCMC samples, while the 2D contours correspond to the 68\% and 95\% regions of a multivariate Gaussian fit to those samples. 
The direct point source measurements are consistent with no sidelobe depolarization ($\betapol^{95}=\betapol^{150}=\betapol^{220}=1$), a conclusion that was disfavored at $5\sigma$ in the C26 power spectrum analysis.}
\label{fig:cosmology_comparison}
\end{figure*}

We compare our direct, point source-based measurements of the optical beam with the constraints previously derived from CMB power spectrum analyses in C26. To perform a rigorous statistical comparison, we extract the publicly available SPT-3G 2019--2020 (``D1'') $\Lambda$CDM chain from \href{https://lambda.gsfc.nasa.gov/}{LAMBDA} and construct the three-dimensional posterior distribution $P(\betapol^{95}, \betapol^{150}, \betapol^{220})$. We fit a trivariate Gaussian to a gridded histogram of the posterior samples, effectively extrapolating beyond the $\betapol \in [0,1]$ prior used in C26 to allow comparison with our point source constraints that extend outside this range. 

Considering the diagonals of the trivariate Gaussian, we find $\betapol = 0.37\pm0.19$ at 95~GHz, $0.49\pm0.27$ at 150~GHz, and $0.55\pm0.21$ at 220~GHz as the 68\% credible intervals from the C26 CMB power spectrum analysis. These central values are lower than the C26 posterior means quoted in Section~\ref{sec:intro} ($0.48$, $0.62$, $0.62$) because the unbounded Gaussian fit used here extends support below the $\betapol \in [0,1]$ prior imposed in C26, pulling the fitted means downward. Comparing these to our Bayesian point-source results (Table~\ref{tab:hyperparameters}), we find that the point source data consistently prefer higher $\betapol$ values, indicating less depolarization than inferred from the power spectrum. To quantify the agreement in three dimensions, we calculate the Mahalanobis distance $\sqrt{(\vec{\mu}_{\mathrm{PS}} - \vec{\mu}_{\mathrm{C26}})^\mathsf{T} (\mathbf{C}_{\mathrm{PS}} + \mathbf{C}_{\mathrm{C26}})^{-1} (\vec{\mu}_{\mathrm{PS}} - \vec{\mu}_{\mathrm{C26}})}$, where $\vec{\mu}_{\mathrm{PS}}$ and $\mathbf{C}_{\mathrm{PS}}$ refer to the parameter vector and covariance from the point source analysis in this work, and $\vec{\mu}_{\mathrm{C26}}$ and $\mathbf{C}_{\mathrm{C26}}$ refer to those from C26. The Mahalanobis norm is $2.1\sigma$, which corresponds for three degrees of freedom to $p=0.21$. We can compare this to an effective one-dimensional disagreement significance of $1.3\sigma$ by calculating at what significance a two-tailed Gaussian distribution has $p=0.21$. This does not meet standard thresholds for claiming statistical inconsistency (e.g., $3\sigma$ or $p < 0.01$).

When measured against only the point source covariance, the C26 best-fit parameters lie at a Mahalanobis distance of $6.9\sigma$, while the point source best-fit values lie at a distance of $2.5\sigma$ from the C26 posterior. For three degrees of freedom, the point source analysis therefore assigns negligible probability to the C26 central value if its uncertainties are ignored. On the other hand, the point-source derived central value is acceptable under the C26 constraints with $p=0.11$. The combined constraint allows a compromise solution with $p=0.21$.

It is important to recall that C26 inferred $\betapol$ indirectly, from the requirement of inter-frequency consistency of the polarization power spectra, whereas the present work measures the polarized beam response directly from point sources. The C26 $\betapol$ parameter could therefore have absorbed several effects other than physical sidelobe depolarization. These include residual band-dependent beam-model inaccuracies, residual temperature-to-polarization leakage, filtering or transfer-function mismatches between observing bands, band-dependent foreground residuals, and other frequency-dependent sources of spurious $E$-mode power. If the $1.3\sigma$ difference is not a statistical fluctuation, any of these effects---rather than true depolarization---could be responsible for the low $\betapol$ values preferred by the power spectrum analysis. Alternatively, the simple single-parameter $\betapol$ sidelobe model may be insufficiently flexible. A more sophisticated beam model that exhibits depolarization at low multipoles, where the power spectrum analysis has greatest sensitivity, while remaining consistent with the temperature beam at the high multipoles probed by point sources, could potentially reconcile the two measurements. We explore this option in Appendix~\ref{app:lrestricted}. We do not claim to identify which of these explanations is correct; we note only that our direct measurement disfavors a simple interpretation in which the sidelobes are strongly depolarized following the C26 sidelobe model. 

Figure~\ref{fig:cosmology_comparison} shows the one- and two-dimensional projections of these three-dimensional constraints. While there is overlap in the 95\% confidence regions, the C26 contours (blue) are systematically offset toward lower values compared to the point source contours (orange). The two methods probe $\betapol$ through entirely different physical effects: our analysis measures the radial beam profile directly from bright sources, while the cosmological analysis infers $\betapol$ through its effect on large-scale CMB power spectra. The direct measurement suggests that the actual optical depolarization is minimal.

To assess whether methodological differences in time constant treatment or data selection contribute to this offset, we performed a retrospective check using only 2019--2020 data processed without time constant deconvolution, matching the methodology of C26. Crucially, this analysis uses the exact same beam profiles $B_\mathrm{main}(r)$ and $B_\mathrm{full}(r)$ used in C26, meaning $\betapol$ implies precisely the same radial profile as in that work (in contrast to our primary analysis, which interpolates between optical beam profiles constructed after time constant deconvolution). We find $\betapol^{95}=0.83\pm0.17$, $\betapol^{150}=0.96\pm0.15$, and $\betapol^{220}=0.61\pm0.35$. 
The discrepancy with the C26 constraint becomes an effective one-dimensional significance of 1.1$\sigma$, down from 1.3$\sigma$ in our primary analysis.
This lack of significant shift supports our expectations that the instrument beam is stable over the 2019--2023 observing period, and that the $\betapol$ parameter provides a self-consistent measure of physical sidelobe depolarization, whether time constant effects are included in the beam model or deconvolved from the TOD.
\\ \\

\section{Discussion}
\label{sec:discussion}

\subsection{Physical Interpretation}

Our direct measurements from polarized point sources yield $\betapol = 0.89 \pm 0.10$ at 95~GHz, $1.08 \pm 0.10$ at 150~GHz, and $0.90 \pm 0.22$ at 220~GHz, consistent with unity across all frequencies. These results indicate that SPT-3G beam sidelobes preserve polarization nearly as efficiently as the main beam, with the data constraining any depolarization to $\lesssim20\%$ at 95 and 150~GHz.

The consistency with $\betapol = 1$ constrains the physical mechanisms that produce beam sidelobes. Processes that destroy polarization information, such as diffuse scattering from rough surfaces or baffling materials, are incompatible with our measurements. Instead, the data favor coherent optical processes that preserve the polarization state of incident radiation. Diffraction at aperture edges and specular reflections from telescope structures are both coherent processes that do not stochastically scramble the polarization signal. Our measurements indicate that such polarization-preserving processes in the SPT-3G optical system are responsible for producing the observed sidelobe response.

\subsection{Implications for Cosmological Analysis}
\label{sec:cosmoimplications}

The difference between our point source measurements and the cosmological constraints of C26 warrants careful interpretation. While the statistical disagreement is only 1.3$\sigma$, it is striking that the direct measurements from point sources point to polarization beams that are entirely consistent with the temperature beams, while C26 rejected this null hypothesis at $5\sigma$.
This observed discrepancy admits three explanations. 
First, it may simply be a statistical fluctuation. 
Second, it may point to a limitation of the beam model itself; the power spectrum constraints are driven by low multipoles ($\ell \lesssim 3000$) while our analysis derives the majority of its constraining power from higher multipoles ($3000 \lesssim \ell \lesssim 10{,}000$), suggesting that a more complex polarized beam profile could reconcile the two measurements. The $\ell$-restricted fits of Appendix~\ref{app:lrestricted} test one such family of models and find no statistically significant scale dependence (the weakest consistency with $\betapol=[1,1,1]$ is $p=0.08$, before accounting for the look-elsewhere effect), but those fits do not span the space of all possible beam models. We have not identified a specific, physically motivated model that would reconcile the two measurements, but we cannot exclude the existence of such a model as the ground truth.
Third, the difference may stem from residual systematics in the power spectrum analysis unrelated to the optical beam being absorbed into the $\betapol$ parameter.

Despite this ambiguity, these measurements provide valuable inputs for future SPT-3G cosmological analyses. Previous analyses demonstrated that marginalizing over $\betapol$ with uninformative priors degrades constraints on cosmological parameters including the Hubble constant, baryon density, and scalar spectral index. 
Our measurements substantially reduce the allowed parameter volume, providing future SPT-3G power spectrum analyses with a $\delta\betapol \sim 0.1$ prior at 95 and 150~GHz, compared to the $[0,1]$ range previously used. 
These informative priors have the potential to sharpen cosmological constraints by breaking the beam-cosmology degeneracies identified in previous works.

The aim of this paper is not to define the full SPT-3G beam error budget or to propagate $\betapol$ through a complete cosmological likelihood. The precise impact on cosmological parameters depends on the full beam covariance, the adopted power-spectrum likelihood, the foreground model, and the calibration treatment, and the complete temperature-beam requirements and uncertainty propagation are addressed in the companion beam paper (H26) and in future power-spectrum analyses. We therefore defer a complete propagation to those analyses, which can use the direct point-source $\betapol$ measurements presented here as external priors. We have not carried out full cosmological simulations varying the mean and width of $\betapol$, as such simulations would require choices about the full power-spectrum likelihood, foreground model, calibration model, and beam-covariance propagation that go beyond the scope of this paper. However, realizing this potential requires determining which of the three explanations above is correct. Resolving whether the observed discrepancy arises from statistical fluctuations, beam modeling limitations, unrelated systematics, or any combination thereof will dictate how these priors are applied in future analyses, and motivates a rigorous search for alternative optical or non-optical systematic explanations for the inter-frequency features observed in the CMB power spectrum.

\section{Conclusions}
\label{sec:conclusions}

We have measured the polarization beam response of SPT-3G using 100 bright polarized point sources observed from 2019 through 2023. The sidelobe polarization efficiency parameter $\betapol$ is constrained to $0.89 \pm 0.10$ at 95~GHz, $1.08 \pm 0.10$ at 150~GHz, and $0.90 \pm 0.22$ at 220~GHz. These measurements are consistent with $\betapol = 1$, supporting identical temperature and polarization beams.

Extensive systematic tests validate these results. The measurements prove stable across alternative statistical formulations (real-space versus Fourier-space), different covariance treatments, variations in leakage template construction, and alternative source selections. The agreement between frequentist bootstrap and Bayesian posterior sampling confirms that uncertainties are well characterized. The B-spline analysis provides complementary highly parametric beam profiles that qualitatively agree between temperature and polarization, further supporting the $\betapol\approx1$ case.

Comparing these results with the $\betapol$ values inferred from previous cosmological power spectrum analyses (C26), we find a $1.3\sigma$ difference. While the power spectrum analysis favored $\betapol \sim 0.5$, our direct measurements favor $\betapol \approx 1$, indicating that the physical beam sidelobes are not significantly depolarized. This difference could stem from residual systematics in the power spectrum analysis unrelated to the optical beam being absorbed into the $\betapol$ parameter. Alternatively, the difference may reflect the distinct angular scales probed by each analysis. The $\betapol$ constraints from the power spectrum are primarily driven by low multipoles ($\ell < 3000$), whereas our point source analysis derives the majority of its constraining power from the range $3000 \lesssim \ell \lesssim 10{,}000$, allowing for a more precise measurement of the shape of $B(\ell)$ at high multipoles. Therefore, the difference in $\betapol$ constraints may point to a limitation of the beam model itself; a more complex polarized beam profile that deviates from the temperature beam only at large angular scales could potentially reconcile the two measurements.

These constraints provide valuable inputs for future SPT-3G cosmological analyses. However, realizing their potential to break the beam-cosmology degeneracies discussed in Section~\ref{sec:cosmoimplications} requires determining which of the three explanations above is correct. By establishing polarization beam properties independently of cosmological modeling, we aim to prevent beam nuisance parameters from absorbing residual systematics or biasing cosmological inferences. Our methodology demonstrates that bright polarized point sources can constrain polarized beam systematics with $\sim10\%$ precision, comparable to the precision achieved through the cosmological analyses themselves.

The computational framework developed here, including GPU-accelerated fitting and modern optimization techniques, provides a template applicable to other high-resolution CMB experiments requiring precise beam characterization. Future SPT configurations, including the planned SPT-3G+ upgrade, will require new polarized beam measurements, and the analysis framework established in this work can be directly applied to characterize these upgraded instruments, ensuring continuity in beam systematics control as the SPT project advances.

\begin{acknowledgments}
The South Pole Telescope program is supported by the National Science Foundation (NSF) through awards OPP-1852617 and OPP-2332483. Partial support is also provided by the Kavli Institute of Cosmological Physics at the University of Chicago. 
This research was done using services provided by the OSG Consortium \citep{pordes07, sfiligoi09, osg06, osg15}, which is supported by the National Science Foundation
awards \#2030508 and \#2323298.
The GPU resources used in this work were provided by the QUP World Premiere Institute. 
Argonne National Laboratory's work was supported by the U.S. Department of Energy, Office of High Energy Physics, under contract DE-AC02-06CH11357. The UC Davis group acknowledges support from Michael and Ester Vaida. Work at the Fermi National Accelerator Laboratory (Fermilab), a U.S. Department of Energy, Office of Science, Office of High Energy Physics HEP User Facility, is managed by Fermi Forward Discovery Group, LLC, acting under Contract No. 89243024CSC000002. The Melbourne authors acknowledge support from the Australian Research Council's Discovery Project scheme (No. DP210102386). The Paris group has received funding from the European Research Council (ERC) under the European Union's Horizon 2020 research and innovation program (grant agreement No. 101001897), and funding from the Centre National d'Etudes Spatiales. The SLAC group is supported in part by the Department of Energy at SLAC National Accelerator Laboratory, under contract DE-AC02-76SF00515. 
\end{acknowledgments}

\appendix

\section{Source-Subset Tests}
\label{app:splitsample}

One possible concern is that the inferred value of $\betapol$ could be driven by a particular subset of the source population. This is especially relevant because the constraining power on the polarized beam is expected to depend strongly on source polarization fraction. We therefore repeat the analysis on two independent halves of the source sample, split by 95~GHz polarization fraction.

The lower-polarization half contains the 50 sources with the smallest 95~GHz polarization fractions. Their mean polarization fractions are 1.0\%, 1.3\%, and 2.4\% at 95, 150, and 220~GHz, respectively. The higher-polarization half contains the remaining 50 sources, with corresponding mean polarization fractions of 3.3\%, 3.6\%, and 4.6\%. For each subset, we repeat the fitting procedure described in Section~\ref{sec:fitting_framework}. Because the subset tests are intended as internal consistency checks rather than as the primary uncertainty estimate, we report conditional constraints on the joint three-band $\betapol$ posterior, with one-dimensional marginalized uncertainties computed analytically from this conditional covariance. These conditional uncertainties are analogous to the Bayesian posterior uncertainties reported in Table~\ref{tab:hyperparameters}, except that the source amplitudes and positions are held fixed at their best-fit values and the $\betapol$ covariance is estimated from the local $\chi^2$ curvature rather than from posterior sampling.

\begin{table}[b]
\centering
\begin{tabular}{l c c c c}
\\
\hline
Analysis & $N_{\rm src}$ & $95\,{\rm GHz}$ & $150\,{\rm GHz}$ & $220\,{\rm GHz}$ \\
\hline
Baseline, bootstrap & 100
& $0.890 \pm 0.098$
& $1.084 \pm 0.095$
& $0.896 \pm 0.219$ \\
All sources, conditional & 100
& $0.890 \pm 0.106$
& $1.084 \pm 0.063$
& $0.896 \pm 0.175$ \\
Least polarized half, conditional & 50
& $0.631 \pm 0.329$
& $1.162 \pm 0.152$
& $1.216 \pm 0.352$ \\
Most polarized half, conditional & 50
& $0.894 \pm 0.108$
& $1.065 \pm 0.068$
& $0.797 \pm 0.197$ \\
\hline
\end{tabular}
\caption{Comparison of the baseline $\betapol$ constraints with conditional constraints from source-subset fits. The baseline row reports the primary bootstrap uncertainty used elsewhere in this work. The remaining rows use the same conditional uncertainty estimate, enabling an apples-to-apples comparison of source subsets split by 95~GHz polarization fraction. The most polarized half of the sample dominates the constraining power, but the least- and most-polarized halves are statistically consistent with each other ($p=0.54$).}
\label{tab:source_split}
\end{table}

The results are summarized in Table~\ref{tab:source_split}. The higher-polarization subset retains most of the constraining power of the full sample, as expected, while the lower-polarization subset gives weaker but statistically compatible constraints. A direct comparison of the two independent subset measurements using the three-dimensional Mahalanobis distance gives $p=0.54$, indicating no evidence for inconsistency between the two halves of the source sample. We therefore retain the full-sample bootstrap result as the baseline measurement, since it uses all available source information and its bootstrap uncertainty estimate already captures source-to-source variation.

\section{\texorpdfstring{$\ell$}{ell}-Restricted \texorpdfstring{$\betapol$}{betapol} Fits}
\label{app:lrestricted}

The comparison between the point source-based measurement and the C26 power-spectrum constraint on $\betapol$ is model-dependent. In the C26 analysis, $\betapol$ was constrained using CMB power spectra over $400 \leq \ell \leq 4000$, whereas the baseline point-source measurement in this work derives a substantial fraction of its constraining power from higher multipoles, extending to $\ell \sim 15{,}000$ (Section~\ref{sec:precision_source}). The two measurements therefore test the same one-parameter beam model over different ranges of angular scale. Since $\betapol$ measures how closely the polarization beam resembles $B_{\mathrm{full}}(r)$ versus $B_{\mathrm{main}}(r)$ (Section~\ref{sec:betapol}), any comparison between the two analyses is conditional on the assumed scale dependence of the difference between these two beam models. To probe this dependence, we repeat the $\betapol$ inference after progressively lowering the maximum multipole included in the point-source fit, thereby removing the high-$\ell$ information that drives much of the baseline constraint. This $\ell$-restricted test assesses whether the inferred polarization beam response remains stable as the point-source analysis is restricted to angular scales more directly comparable to those used in C26, and whether the apparent agreement or disagreement depends on the adopted $\left( B_{\mathrm{full}}-B_{\mathrm{main}} \right)$ sidelobe model.

Because this test requires repeated fits of the full model, we report conditional rather than full bootstrap uncertainties for the $\ell$-restricted results. For each value of $\ell_{\mathrm{max}}$, we refit the model and compute the conditional covariance of the three-band parameter vector $[\betapol^{95},\betapol^{150},\betapol^{220}]$ from the local curvature of the $\chi^2$ surface. The quoted one-dimensional uncertainties are then obtained by analytically marginalizing this three-dimensional Gaussian approximation over the other two frequency bands. These uncertainties are intended for internal comparison across $\ell_{\mathrm{max}}$ cuts, while the baseline result and its bootstrap uncertainty remain the primary measurement.

\begin{figure}
\centering
\includegraphics[width=\linewidth]{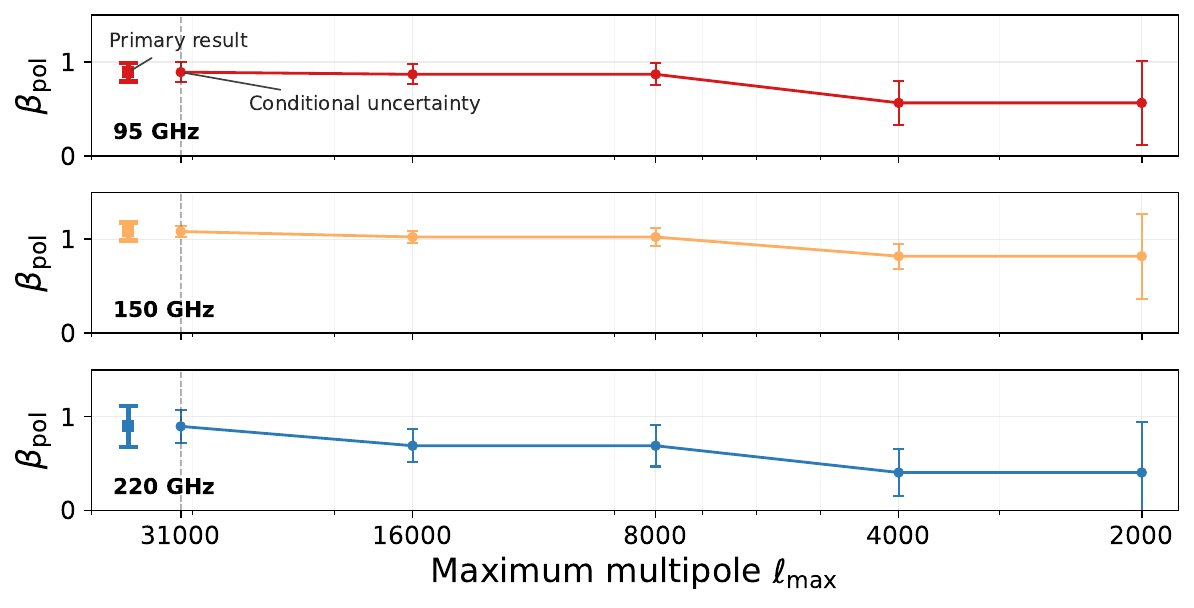}
\caption{$\ell$-restricted $\betapol$ constraints as a function of the maximum multipole included in the fit. For each value of $\ell_{\mathrm{max}}$, the analysis is restricted to modes with $\ell \leq \ell_{\mathrm{max}}$ and the $\betapol$ parameters are refit. Thin error bars show conditional one-dimensional uncertainties derived from the curvature of the three-dimensional $\chi^2$ surface in $[\betapol^{95},\betapol^{150},\betapol^{220}]$, after analytic marginalization over the other two frequency bands. Thick error bars show the baseline bootstrap result for comparison. The best-fit values show a mild tendency toward lower $\betapol$ as $\ell_{\mathrm{max}}$ is reduced, but the trend is not statistically significant.}
\label{fig:ellmax}
\end{figure}

The results are shown in Figure~\ref{fig:ellmax}. As $\ell_{\mathrm{max}}$ is lowered, the best-fit values show a mild tendency toward smaller $\betapol$. However, this trend is not statistically significant. The model with identical normalized temperature and polarization beams, $\betapol=[1,1,1]$, remains compatible with the data for all $\ell_{\mathrm{max}}$ cuts considered. The smallest consistency probability occurs for $\ell_{\mathrm{max}}=4000$, where $\betapol=[1,1,1]$ is still allowed at $p=0.08$ before accounting for the look-elsewhere effect from examining multiple cuts.

We therefore find no statistically significant evidence, at the present point-source signal-to-noise, for scale-dependent behavior inconsistent with the $\betapol$ model. At the same time, the restricted-$\ell$ fits demonstrate that the $\simeq\!10\%$ precision of the baseline measurement relies on the higher-$\ell$ information available in the point-source maps. At lower multipoles, where the comparison to the C26 power-spectrum analysis is more direct, the point-source constraints are substantially weaker.

This test therefore does not identify scale dependence as the explanation for the C26 preference for lower $\betapol$. It also does not rule out more general scale-dependent polarized-beam structure outside the one-parameter $\betapol$ model family, such as a partially depolarized component localized to angular scales where the present point-source data have limited constraining power.

\section{Analysis Improvements and Pipeline Validation}
\label{app:postsub}

During development and validation of the analysis pipeline, we identified and corrected several issues. Each correction had a modest effect on the inferred values of $\betapol$, and the results reported throughout this work incorporate their combined effect. All figures were generated using the validated analysis code. We summarize the relevant corrections and their approximate effects on $\betapol$.

\begin{itemize}
\item \textbf{Main-beam $r=0$ extrapolation (${\sim}0.1\sigma$).} The physical-optics model for $B_{\mathrm{main}}(r)$ contained an error in the treatment of the central beam value. Rather than evaluating the response directly at $r=0$, the previous implementation extrapolated this point from the first finite-radius sample. This produced a small, unphysical flattening of the beam core. Correcting this evaluation removes the spurious core feature and slightly shifts the inferred $\betapol$ values.

\item \textbf{Treatment of Saturn data (${\sim}0.2\sigma$).} The initial analysis combined the point-source-derived near beam with Saturn-derived far-sidelobe measurements using a one-dimensional stitching procedure. This procedure required band-dependent choices for the transition radius and relative normalization between the two data sets, and the resulting $\betapol$ constraints showed non-negligible sensitivity to these choices. In the updated analysis, the point-source data are used directly out to the $0.1\%$ beam level, and the Saturn far-sidelobe measurements are incorporated through a joint likelihood rather than through a fixed radial stitching prescription. Across a broad range of reasonable hyperparameter choices, this updated treatment changes the downstream $\betapol$ constraints by less than $0.05\sigma$, indicating substantially improved robustness.

\item \textbf{CMB covariance sign convention (${\sim}0.2\sigma$).} The Fourier-domain covariance model previously transformed the theoretical CMB $TE$ spectrum to the SPT-3G map-frame Stokes basis using an incorrect sign convention for $U$. In the IAU convention, the polarization angle is defined in a local basis whose $x$-axis points North and whose $y$-axis points East. Since the SPT-3G map coordinates do not use this basis, the corresponding Fourier-space angle is $\phi_{\rm IAU}=\mathrm{arctan2}(-k_x,k_y)$. The previous implementation used the opposite sign convention for this transformation, flipping the signs of the $TU$ and $QU$ covariance terms and slightly misestimating the Fourier-domain precision matrix. This error does not produce an ensemble-mean bias in the fitted beam parameters: applying an incorrect but fixed weighting to an otherwise correctly specified mean model changes the estimator variance but not its expectation value. For the particular data realization analyzed here, however, the modified weighting of Fourier modes produces a small shift in the recovered parameters.

\item \textbf{Apodization window function (${\sim}0.3\sigma$).} We identified an error in the window function used to sample the CMB power spectra when constructing the Fourier-domain precision array. The relevant window function is the Fourier transform of the apodization mask, a $300\times300$ pixel map with a 10-pixel raised half-cosine taper at each edge. The precision array is evaluated on a Fourier grid truncated to $\ell_{\mathrm{max}}$, corresponding to an $87\times87$ grid for this analysis. In the previous implementation, the 10-pixel taper width from the full-resolution map was retained when constructing the truncated-grid window function, exaggerating the effect of the window function on the precision matrix. The updated implementation constructs the window function consistently on the truncated Fourier grid.

\item \textbf{Maximum allowed correlations (${\sim}0.2\sigma$).} In constructing the precision matrix, we allow correlations between observing bands and Stokes parameters. In the initial analysis, these correlations were clipped to the range $[-0.95,0.95]$. Such large correlations allow strong cancellations between modes and can therefore make the result sensitive to small inaccuracies in the modeled sky covariance, window-function treatment, or data-driven covariance estimate. This is especially relevant for modes where the empirical covariance estimate fluctuates low because of the finite number of samples. We therefore adopt a more conservative clipping range of $[-0.80,0.80]$ in the updated analysis.
\end{itemize}

Taken together, these corrections have their largest cumulative effect at 150~GHz, while the main-beam $r=0$ extrapolation and CMB covariance sign convention have a negligible net effect on $\betapol^{150}$. Given the statistical uncertainty of 0.10, the result remains consistent with $\betapol=1$. We adopt the corrected analysis because the changes described above improve the physical fidelity and robustness of the pipeline.

\bibliographystyle{aasjournalv7}
\bibliography{spt,betapol}

\end{document}